\documentclass[sigconf]{acmart}

\AtBeginDocument{%
  }

\copyrightyear{2025}
\acmYear{2025}

\setcctype{by}
\acmConference[CHI '25]{CHI Conference on Human Factors in Computing Systems}{April 26-May 1, 2025}{Yokohama, Japan}
\acmBooktitle{CHI Conference on Human Factors in Computing Systems (CHI '25), April 26-May 1, 2025, Yokohama, Japan}\acmDOI{10.1145/3706598.3714088}
\acmISBN{979-8-4007-1394-1/25/04}

\newcommand{\sruthi}[1]{{\color{black} {#1}}}

\usepackage{colortbl}
\usepackage{float}  
\usepackage{subfigure}
\usepackage{multirow}

\begin{document}

\title{The Interaction Layer: An Exploration for Co-Designing User-LLM Interactions in Parental Wellbeing Support Systems}


\author{Sruthi Viswanathan}
\email{sruthi.viswanathan@cs.ox.ac.uk}
\orcid{0000-0002-1113-7171}
\affiliation{%
  \institution{University of Oxford}
  \city{Oxford}
  \country{United Kingdom}
}

\author{Seray Ibrahim}
\email{seray.ibrahim@kcl.ac.uk }
\orcid{0000-0001-9358-6802}
\affiliation{%
  \institution{Kings College London}
  \city{London}
  \country{United Kingdom}
}

\author{Ravi Shankar}
\email{ravi.shankar@wrh.ox.ac.uk}
\orcid{0009-0006-1778-873X}
\affiliation{%
  \institution{University of Oxford}
  \city{Oxford}
  \country{United Kingdom}
}

\author{Reuben Binns}
\email{reuben.binns@cs.ox.ac.uk}
\orcid{0000-0002-8272-5667}
\affiliation{%
    \institution{University of Oxford}
    \city{Oxford}
    \country{United Kingdom}
}

\author{Max Van Kleek}
\email{max.van.kleek@cs.ox.ac.uk}
\orcid{0000-0003-3873-6366}
\affiliation{%
    \institution{University of Oxford}
    \city{Oxford}
    \country{United Kingdom}
}

\author{Petr Slovak}
\email{petr.slovak@kcl.ac.uk }
\orcid{0000-0001-8458-7715}
\affiliation{%
  \institution{Kings College London}
  \city{London}
  \country{United Kingdom}
}







\renewcommand{\shortauthors}{Viswanathan et al.}

\begin{abstract}
Parenting brings emotional and physical challenges, from balancing work, childcare, and finances to coping with exhaustion and limited personal time. Yet, one in three parents never seek support. AI systems potentially offer stigma-free, accessible, and affordable solutions. Yet, user adoption often fails due to issues with explainability and reliability. To see if these issues could be solved using a co-design approach, we developed and tested NurtureBot, a wellbeing support assistant for new parents. 32 parents co-designed the system through Asynchronous Remote Communities method, identifying the key challenge as achieving a ``successful chat.'' As part of co-design, parents role-played as NurtureBot, rewriting its dialogues to improve user understanding, control, and outcomes. The refined prototype, featuring an \textit{Interaction Layer}, was evaluated by 32 initial and 46 new parents, showing improved user experience and usability, with final CUQ score of 91.3/100, demonstrating successful interaction patterns. Our process revealed useful interaction design lessons for effective AI parenting support.
\end{abstract}


\begin{teaserfigure}
  \includegraphics[width=\textwidth]{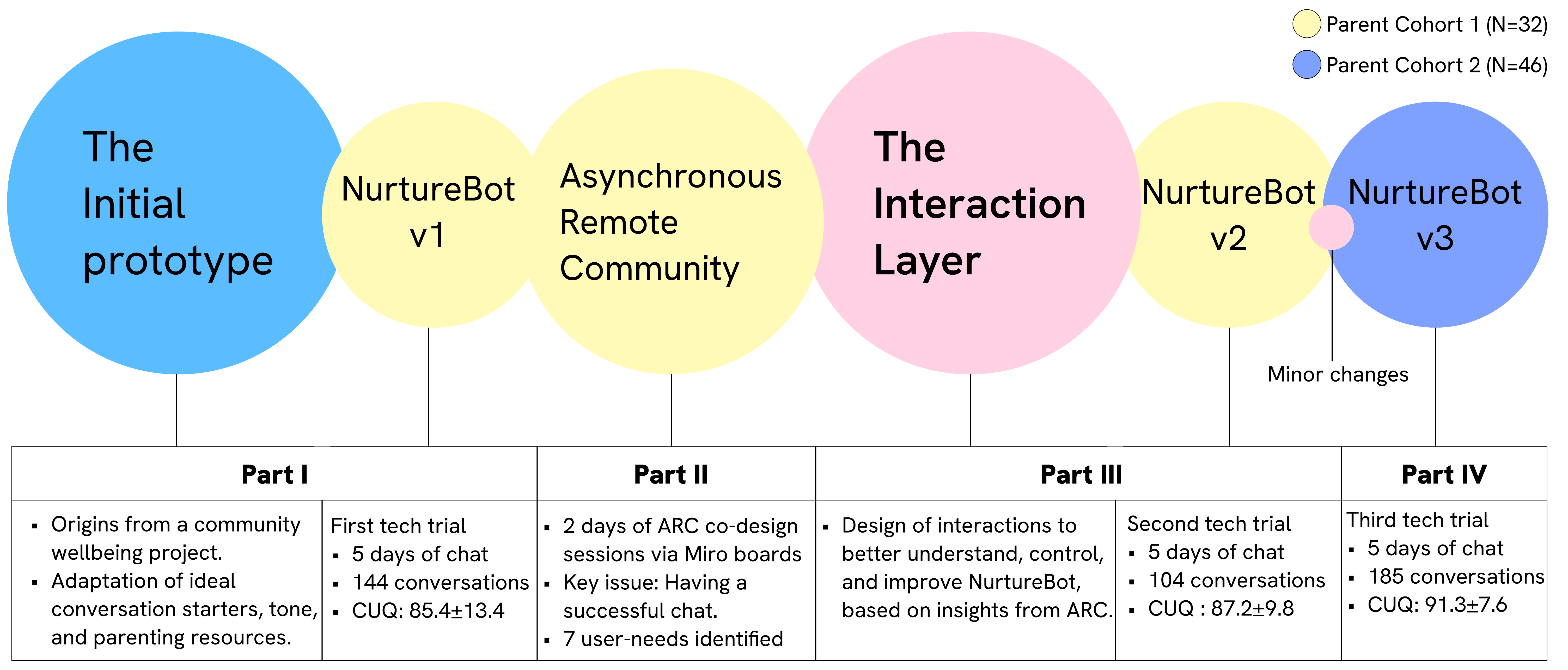}
  \caption{Overview of the four-part study detailing the iterative development and testing of NurtureBot, a parental wellbeing assistant.}
  \Description{A visual representation of the four-part iterative development of NurtureBot involving two parent cohorts. The chart shows the progression from the initial prototype to NurtureBot v1, followed by ARC co-design sessions, NurtureBot v2, and NurtureBot v3. Key findings include conversation starters, tone, parenting resources, and interaction improvements. NurtureBot v1 identified seven major problems, leading to an interaction layer for NurtureBot v2, which addressed four feature requests. NurtureBot v3 showed further improvements with five new feature requests and achieved the highest CUQ score of 91.3.}
  \label{fig:teaser}
\end{teaserfigure}

\maketitle

\section{Introduction}
Parenting, often hailed as the most important job in our world, exacts a significant toll on both mind and body. The perinatal period represents a critical developmental stage for the child, the parent(s), and the formation of attachment relationships \cite{li2017social, allan2013perinatal}. This period, spanning from the onset of pregnancy to the first year postpartum, is marked by profound physical, emotional, and psychological changes, which significantly impact parental wellbeing. In particular, many expectant and new mothers encounter substantial challenges, in mental health, emotional adjustment, and caregiving \cite{widarsson2012support}, in addition to the transformation their body goes through. During this period, parents experience heightened stress, anxiety, and feelings of isolation, but only one in three parents seek help \cite{unicef2022}. Perinatal care literature identifies several key issues, including inadequate access to mental health resources, availability of healthcare professionals, cost, stigma surrounding parental mental health, and challenges in identifying and supporting at-risk mothers \cite{rallis2014transition, davis2022understanding, wieck2017risks}. These challenges have been further aggravated by the COVID-19 pandemic \cite{bridle2022supporting, kinser2022s}. As a result, we see a growing body of research exploring how digital technologies can enhance perinatal care \cite{van2018ehealth,ginja2018associations}.

Notable Human-Computer Interaction (HCI) research has laid the groundwork for exploring how digital interventions can increase wellbeing across aspects such as physical and mental health of a family \cite{monitorhealthjulie}, a child's growth \cite{babysteps}, sleep tracking of parents and children \cite{dreamcatcher}, and managing family screen-time behaviours \cite{screen-media}. Today, with the availability of Artificial Intelligence (AI) to everyday users, the use of general generative applications such as ChatGPT for parenting is on the rise \cite{chatGPTParenting}. With responsible design, Large Language Models (LLMs) have the potential to facilitate a range of services, from offering self-guided mental health support \cite{sharma2024facilitating}, real-time emotional support \cite{kang2024can} to providing evidence-based information at scale \cite{lai2023psy}. The current use of LLMs to support parental and child wellbeing raises both optimism, but also ethical caution \cite{ashraf2024systematic, stade2024large}, due to limitations and uncertainties regarding accuracy of information provided, empathy, and usefulness \cite{kaneda2023artificial}.

Given these challenges, our research adopts an exploratory approach, probing into a small subset of these broader issues. We focus on understanding how interactions between LLMs and parents might be designed to be helpful and engaging. Our goal is to explore what constitutes a usable and useful interaction in these contexts and how LLMs can be integrated into wellbeing support in ways that are practically engaging and thus useful for parents. To this end, we conducted a four-part online study involving 78 parents, who trialled iterative improvements of \textit{NurtureBot}, a technology probe, designed to support parental wellbeing. This study, while not aimed at delivering a finalised solution, served as an exploratory step toward understanding how parents interact with LLM-based agents and identifying necessary features to make these tools genuinely supportive. The following Research Questions (RQs) guide our investigation into the user interaction aspects of engaging with an LLM-based parental wellbeing assistant:
\begin{itemize}
    \item \textbf{RQ1:} What are the main problems parents face when interacting with an LLM-based parental wellbeing assistant in support-seeking scenarios?
    \item \textbf{RQ2:} What key user needs must a parental wellbeing assistant address? How do parents describe their ideal vision of how the assistant should function, particularly regarding their understanding of its workings, control over it, and opportunities to improve the outcomes?
    \item \textbf{RQ3:} How can we employ co-design methods to craft interactions around AI, enhancing usability and the overall user experience? 
\end{itemize}

\paragraph{Study Flow Overview} This four-part user-centred design process, as illustrated in Figure \ref{fig:teaser}, guided the iterative development of NurtureBot. To mitigate emotional risks, we engaged a `slightly-out-of-target' group of parents \cite{Islind2023ProxyDesign}, each with a child aged five or six. These parents simulated stages of their perinatal period and reflected on past challenges with NurtureBot. Using a mixed-methods approach, we gathered both qualitative and quantitative feedback by testing NurtureBot v1, v2, and v3, each over a five-day period. On the final day of each testing period, participants completed the Chatbot Usability Questionnaire (CUQ) \cite{cuq}, assessing usability and engagement (see Figure \ref{survey}).
\begin{itemize}

    \item \textbf{Part I: The Initial Prototype: NurtureBot v1} The development of NurtureBot v1 leveraged LLM technology, informed by a previous community wellbeing initiative run by Committee For Children \cite{CommitteeForChildren} to support new parents. Zero-shot prompting was used to integrate supportive conversation starters and key features, including empathetic chatting, wellbeing exercises, and parenting information. A cohort of 32 parents engaged with NurtureBot v1, generating 144 conversations. Thematic analysis identified seven key issues faced by the user. NurtureBot v1 achieved a CUQ score of 85.4, surpassing the benchmark of 68 \cite{cuq}, indicating good usability.

    \item \textbf{Part II: ARC Co-Design Study} The Asynchronous Remote Community (ARC) \cite{ARC,arcmothers} co-design sessions engaged the same cohort of parents with Miro board activities \cite{miro2011}. Parents prioritised the key issue as having a ``successful chat,'' and reimagined and rewrote NurtureBot's dialogues to improve user understanding, control, and outcomes; thematic analysis of revised dialogues yielded seven refined user needs. 
    
    \item \textbf{Part III: The Interaction Layer: NurtureBot v2} NurtureBot v2 introduced an \textit{interaction layer} with four essential components (See Figure \ref{prompt}): (i) A core interaction principle that enabled users to "Understand, Control, and Improve" exchanges, (ii) Interaction states that directed the chatbot’s behaviour across various tasks initiated by the user, (iii) Interaction levels based on few-shot prompting, incorporating co-designed dialogues from ARC sessions, and (iv) Interaction elements, tailored to address user needs identified during ARC. NurtureBot v2 underwent testing with the same initial cohort, yielding 104 conversations, revealing some remaining minor bugs, while achieving a CUQ score of 87.4, indicating further improvement.

    \item \textbf{Part IV: Prototype Improvement: NurtureBot v3 and Validation} To ensure an unbiased evaluation, we refined NurtureBot further and tested v3 with a new cohort of 46 parents. Feedback analysis identified two new problems to be considered for future iterations, and a CUQ score of 91.3, marking a robust improvement in usability.
    
\end{itemize}

Our contributions are:
\begin{itemize}
    \item \textbf{Empirical Findings and Implications} for co-designing `usable and useful' LLM-based parental wellbeing support systems, from a longitudinal four-week iterative study, gathered from 78 parents, eliciting the interactional challenges they encountered and the solutions proposed to address these issues.
    \item \textbf{The Interaction Layer} a prompt architecture that affords in-situ user understanding, user control, and conversational improvements by directly incorporating end-user inputs as prompt instructions usable by LLMs, addressing usability challenges to enhance user experience with AI-driven support tools for parenting and beyond.

\end{itemize}

This study initiates our exploration into co-designing agents for parental wellbeing, setting foundations for future research to refine and expand AI tools. The diverse parenting experiences and emotional complexities highlight the need for continued development and testing.

\section{Related Work}

\subsection{Supporting Family Wellbeing with Digital Tools}

Within the field of HCI, there has been a long-term research agenda focusing on the study and development of impactful digital tools that offer a variety of social, physical, and mental wellbeing support to parents and mothers specifically \cite{d2016feminist, thomas2019technology, britton2019mothers}. The design and evaluation of digital tools aimed at promoting maternal wellbeing during the transition to motherhood \cite{newhouse2019}, \textit{bump2bump}, emphasise the need for holistic approaches in HCI that consider the challenges faced by new mothers. The role of peer support in digital environments is highlighted by the finding that online networks can provide valuable emotional and practical assistance to parents, particularly in situations where traditional face-to-face support may not be available \cite{yamashita2022online}. While apps supporting first-time mothers, such as \textit{Baby Buddy} in the UK \cite{drakatos2024}, are generally well-received, effectiveness depends on alignment between usability and the mothers' understanding of their needs, suggesting a need to involve parents in the design process to ensure that digital tools provide meaningful support.

The integration of AI chatbots into parenting and family well-being has opened new avenues in this context \cite{yu2023,entenberg2021, wong2021chatbot}. Recent studies have focused on parenting with conversational agents such as Alexa \cite{alexaparenting}, Siri \cite{sezgin2022hey} and other voice-activated agents in the home \cite{sun2021child}. The integration of LLMs into parenting practices \cite{chatGPTParenting}, presents new possibilities and challenges, with LLM-based agents such as ChatGPT being used by parents of young children \cite{chatGPTParenting, kaneda2023artificial}, though their impact on parenting remains largely unexplored. The performance and acceptance of ChatGPT's responses in the childcare field \cite{kaneda2023artificial}, is also being studied with caution. Despite technological advancements, several limitations persist such as misinformation \cite{kaneda2023artificial}, lack of diversity \cite{bajwa2024}, and threat of data misuse \cite{galea2023}. While gamification can be an effective motivational tool \cite{wernbacher2022}, it does not appeal to all users equally. The need for humane interaction with a good user experience has been repeatedly reported in the literature for digital health tools built with or without AI \cite{bautista2023understanding}. A review of 13 AI-based mental health apps with user ratings over 4 out of 5, revealed limited support for AI literacy and explainability \cite{alotaibi2023review}. A study of engagement with 93 health apps indicates that while install rates are high, sustained user engagement remains low, with the median 15- and 30-day retention a mere 3.9\% and 3.3\% respectively \cite{baumel2019objective}. These gaps suggest a need for further research and development of applications which can deliver sound AI-powered digital tools for wellbeing in a sustained way.

\subsection{Interacting with AI for Everyday Users}
\label{HCAI}


As AI permeates daily life, from recommendation systems to digital assistants, the ability of non-expert users to engage meaningfully with AI systems becomes critical \cite{eiband}. While Explainable AI (XAI)  tools enable expert users to grasp the rationale behind AI outputs,  \cite{gunning2019xai, miller2019explanation, carvalho2019machine}, most non-experts do not engage with XAI \cite{miller2023explainableaideadlong}, highlighting the need for innovative methods to foster AI understanding \cite{miller2023explainableaideadlong}. Control over AI systems by everyday users is another essential dimension of Human-Centred AI (HCAI), with research underscoring the importance of user agency in human-AI interactions and challenges with it \cite{konig2024challenges}. Interactive interfaces that allow users to refine AI behaviour, such as setting parameters, have been shown to significantly enhance user satisfaction and the alignment of AI outputs with user expectations \cite{genni_ritter2021enhancing, amershi2014power, holstein2019improving, sitRec}, maintaining user engagement, and ensuring that AI systems serve the intended purposes \cite{bruhlmann2020trust, cai2019hello}. Improvement of AI systems through user feedback represents the third crucial dimension of HCAI. Involving end-users in the process, is not only about improving the AI but also about ensuring that the systems evolve in ways that are aligned with user needs and ethical considerations \cite{reich2019ml}. In summary, HCAI efforts \cite{capel} can be categorised into three fundamental objectives: understanding, controlling, and improving AI systems. Borrowing from the cybernetic loop \cite{patten1981cybernetic}, akin to sensing, processing, and reacting; these categories form the basis for a continuous cycle of interaction between humans and algorithms. We plan to employ these three lenses as cues for our co-design of user-LLM interactions, ensuring that the development of AI systems remains aligned with the needs and expectations of everyday users \cite{cybernetics_weiner1948cybernetics, xu2019perspectives}.

The design of Conversational User Interfaces \cite{landay2019conversational, diederich2022design, clark2019challenges} requires a deep understanding of HCI principles, ensuring that these systems can handle the nuances of human dialogue, such as turn-taking, repair mechanisms, and emotional engagement \cite{landay2019conversational, cruz2019therapeutic, kaye2018voice}. As the field progresses, there is a growing emphasis on addressing these challenges to enhance the user experience and make conversational interactions more natural and effective \cite{lee2019accelerating, song2019interpersonal}.

\sruthi{
The advent of LLMs has propelled CUIs into a wide range of domains. In healthcare, LLMs have been integrated into administrative and clinical tasks, purporting to streamline operations such as patient record management, appointment scheduling, and even aiding in medical diagnostics, in the name of enhancing efficiency and reducing the burden on healthcare professionals \cite{gebreab2024llm, xu2024llm, brown2020language}. Another potential application of LLMs is in personal wellbeing, with agents offering personalised interactions that aim to help users manage stress, anxiety, and other mental health conditions \cite{ferrara2022empowering,ke2024exploringfrontiersllmspsychological}. A systematic review \cite{abd-alrazaq2020effectiveness} of chatbots in improving mental health outcomes, concluded that chatbots might effectively reduce symptoms of depression, anxiety, and stress, while enhancing well-being and quality of life. However, the authors emphasised the necessity for rigorous evaluation of these tools to ensure their safety and efficacy.

\textit{ChaCha} \cite{seo2024chacha}, a chatbot leveraging LLMs was developed to encourage children to share emotions related to personal events, demonstrating the potential of LLMs in facilitating emotional expression among younger populations. Key findings underline the chatbot's ability to facilitate empathetic, developmentally appropriate interactions, though challenges such as balancing engagement and over-reliance were noted. Similarly, \textit{MindfulDiary} \cite{kim2024mindfuldiary}, an LLM-powered tool was designed to assist psychiatric patients in journaling, was tested in a four-week field study involving 28 patients and five psychiatrists, demonstrating that the app facilitated consistent journaling, aided self-reflection, and enhanced patient-clinician communication. This study highlights the potential of LLMs to bridge gaps in mental health care, while addressing challenges of engagement, safety, and integration into clinical workflows. Another recent work explored using LLMs to craft narratives about postpartum depression \cite{postpartumllm}, generating 45 stories based on themes from online support communities. 85\% of the generated narratives successfully adhered to prompts, demonstrating a potential for addressing perinatal mental health challenges through personalised storytelling. However, the stories were also reported to frequently fall short of capturing the emotional depth and relatability inherent in user-generated stories, leaning instead towards methodical and structured exposition.

Despite the promising advancements in LLMs within healthcare, issues such as hallucinations \cite{xu2024hallucination}, task-specific limitations \cite{mizrahi2024state}, and user experience problems persist \cite{wang2024understandinguserexperiencelarge}. The ethical considerations surrounding AI-driven wellness applications have been scrutinized, particularly regarding their potential health risks and regulatory challenges \cite{naturemedicine2024healthrisks}. While the studies described above suggest that generative AI chatbots can offer meaningful assistance to users in the wellbeing domain, concerns persist about their safety and effectiveness, necessitating further research to validate their use in unprompted, unguided real-world scenarios \cite{nature2024generativeai}. Adopting a more human-centred approach may help to build reliable, interactive and creative agents tailored to unique needs \cite{shankar2024validates}. In the context of parenting + AI, this calls for a closer examination of how AI can offer useful and usable support while addressing the unique challenges in this domain.}

\section{Part I: The Initial Prototype: NurtureBot v1}
\label{part1}
\subsection{Origins and motivations for the initial prototype}
Our parental wellbeing support prototype was built upon Nurture, a community wellbeing project originally run by the nonprofit organisation Committee for Children (CFC) \cite{CommitteeForChildren}, which is dedicated to promoting children's wellbeing. The original Nurture project was a peer-to-peer mentoring initiative designed to support new mothers during their baby's first year. In this program, experienced volunteer mothers were trained to mentor new moms, providing guidance and encouragement. A digital platform supporting this program offered personalised, week-by-week content delivered via text messages, acting as a conversation catalyst between mentors and mentees. These messages were designed to initiate meaningful dialogue and help new moms navigate early motherhood. 


The Nurture project was conducted over a span of three years, involving approximately 60 mentors and 603 parents who participated in one-on-one texting. While the project was successful in providing personalised support and the fostering relationships through these interactions, this model faced scalability challenges. While effective, Nurture project was resource intensive and difficult to expand to a broader audience without significant investment in human resources.

\subsection{Designing the initial prototype}
Recognising the need for a scalable solution that could retain the core support offered by Nurture, we partnered with CFC to collate resources and findings from Nurture project for developing a new AI tool. Based on our learning from the Nurture project, we developed an initial prompt for a basic LLM-based chatbot by zero-shot prompting Open AI's GPT-4 API \cite{OpenAI2023GPT4}, which we named NurtureBot v1. We then refined NurtureBot v1's prompt architecture and performed red-team testing \cite{hong2024curiositydrivenredteaminglargelanguage}, by involving the two project managers of the original Nurture project, who had previously designed, facilitated, and trained the volunteer moms. Our research team then independently crafted and iteratively refined the prompt.

NurtureBot v1 was designed to replicate the supportive environment of the original peer-to-peer texting interactions while introducing the scalability and accessibility of AI. The chatbot was refined to deliver \textbf{three key features}: empathetic chatting, wellbeing exercises, and parenting information, all adapted for AI-driven interactions. Additionally, it incorporated ``conversation starters'' thoughtfully crafted by Nurture mentors to begin with and was prompted to perform ``contextual reasoning'' of the user's situation before making suggestions. \textbf{Empathetic chatting} was intended to provide `active listening' that reflects warmth and understanding. \textbf{Wellbeing exercises} were included to offer practical tools for managing stress and mental wellbeing, similar to those shared by mentors. Lastly, \textbf{parenting information} was integrated to ensure that users could access reliable and evidence-based resources tailored to their needs, much like the information shared in the original peer support sessions. \sruthi{It is important to note that the development of NurtureBot v1 was limited to zero-shot prompting of the pre-trained GPT-4 model (see Figure \ref{promptarch}) and a database of curated resources from the Nurture project to guide its responses. No additional training or fine-tuning of the model was conducted, as we intentionally chose not to train the chatbot on human-to-human conversations, which were deeply rooted in the development of interpersonal relationships and emotional bonding between mentors and new mothers.}


\subsection{NurtureBot v1 Testing and Analysis}
\label{part2}
In this phase, we address RQ1: \textit{``What are the main problems parents face when interacting with an LLM-based parental wellbeing assistant in support-seeking scenarios?''}.

\subsubsection{Participants and Method}
\label{cohort1}
Once we had secured ethical approval via our institutional ethics board [King's College London Ethics review - MRA-23/24-44973], we recruited 32 parents of children aged 5 or 6 through the Prolific platform \cite{prolific}. The average age of participants was 41.17 years, with 20 females and 12 males. Five participants (15\%) were in the habit of using ChatGPT everyday. Participants self-reported no acute stress or mental or physical health issues during the study period, nor to have received therapy in the 12 months prior. The study’s purpose and methodology were explained to parents through an information sheet, after which informed consent was obtained. They were paid ~\$3 for each day of participation.

The decision to include parents of children aged 5 or 6 as participants in our study was a carefully considered methodological choice aimed at balancing ethical considerations and the practicalities of our exploratory work. We consciously selected this `slightly out-of-target' group as they are likely to still vividly recall and relate to challenges from early parenthood, yet are past the most immediate and demanding stages of infant care. This choice minimised potential emotional risks, as participants in the acute perinatal period might find discussions about parenting challenges triggering without adequate real-time support safeguards. Moreover, our approach ensured that the study was not overly time-intensive for participants already navigating complex, daily baby-related routines. In addition, leveraging adjacent populations is an established approach in exploratory and design studies, similar to engaging individuals with prior therapy experience to inform therapeutic AI systems \cite{veldmeijer2023involvement}. Our research aimed to simulate and reflect upon perinatal experiences rather than measure clinical outcomes such as Generalised Anxiety Disorder (GAD), aligning it with ethical AI principles that discourage unnecessary exposure of vulnerable populations to experimental settings. By recruiting participants with lived perinatal experience, we gained valuable insights while safeguarding the wellbeing of a demographic closely related to our target population. This approach aligns with exploratory HCI methodologies such as proxy design \cite{Islind2023ProxyDesign} and ethical guidelines outlined in digital parenting support studies \cite{bera_ethics_guide}.

Over the five-day period, participants engaged with NurtureBot v1 for at least five minutes each day, with each day dedicated to simulating a specific stage of early parenthood: pregnancy, the week of childbirth, few weeks postpartum, one month, and six months after birth. The objective was to assess how well the NurtureBot v1 could support them in these hypothetical scenarios. \sruthi{Studies on well-being chatbots support the feasibility of short, focused interactions. For example, the 21-Day Stress Detox chatbot demonstrated effectiveness with sessions lasting approximately 5–7 minutes, providing stress management techniques and maintaining user engagement \cite{stressdetox2021}. Similarly, a recent study on GPT-3-based mental well-being chatbots designed five-minute interactions to help users manage their mood, further validating the efficacy of short interaction durations in achieving meaningful outcomes \cite{gpt3wellbeingchatbot2022}. These examples align with our approach of employing five-minute everyday interactions for five days to capture user insights effectively.}

\begin{figure}[t]
  \centering
  \includegraphics[width=\linewidth]{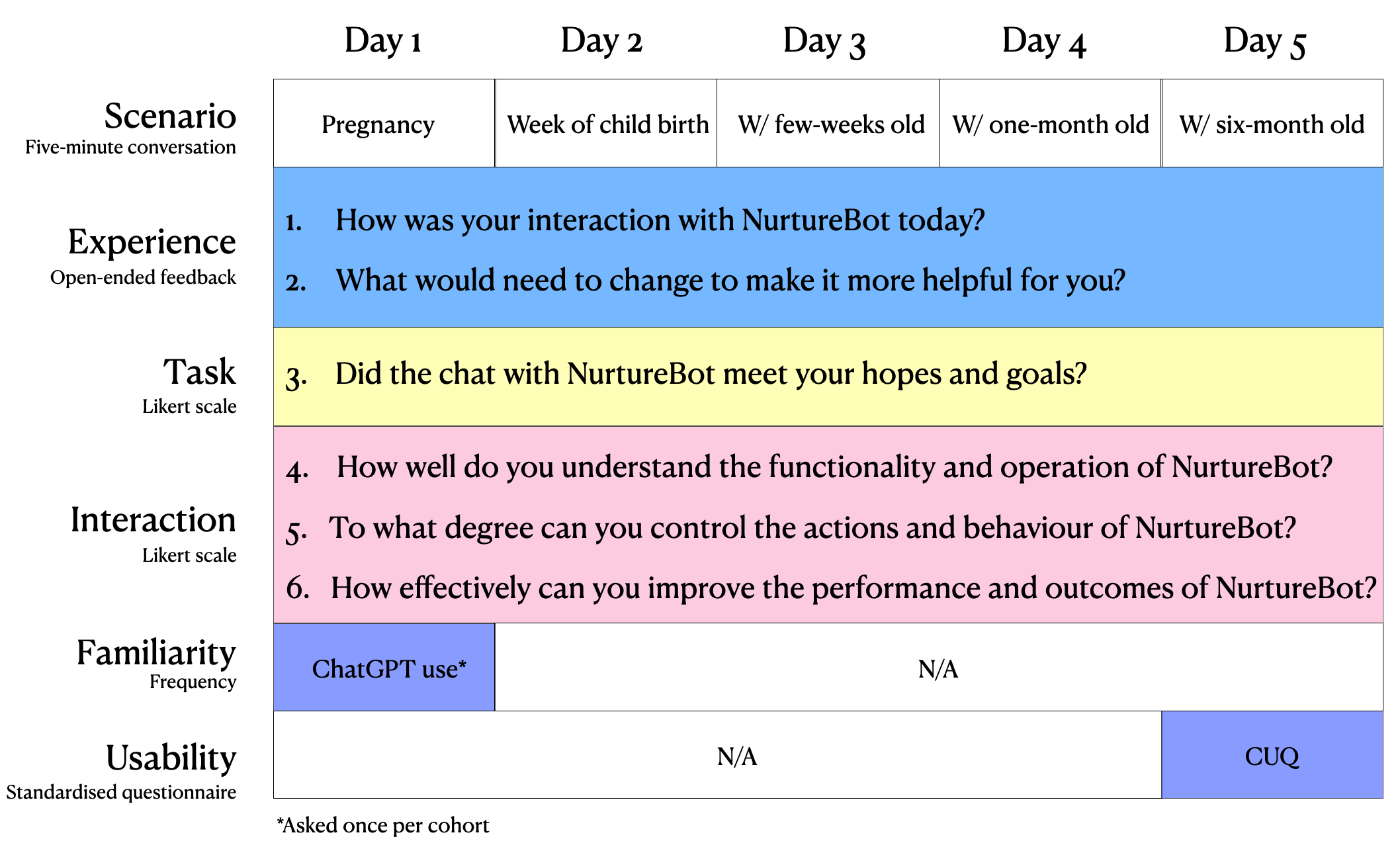}
  \caption{Survey Questions for Each Day in NurtureBot v1, v2, and v3 technology trials}
  \Description{xxx}
  \label{survey}
\end{figure}

\begin{figure}[t]
  \centering
  \includegraphics[width=250px]{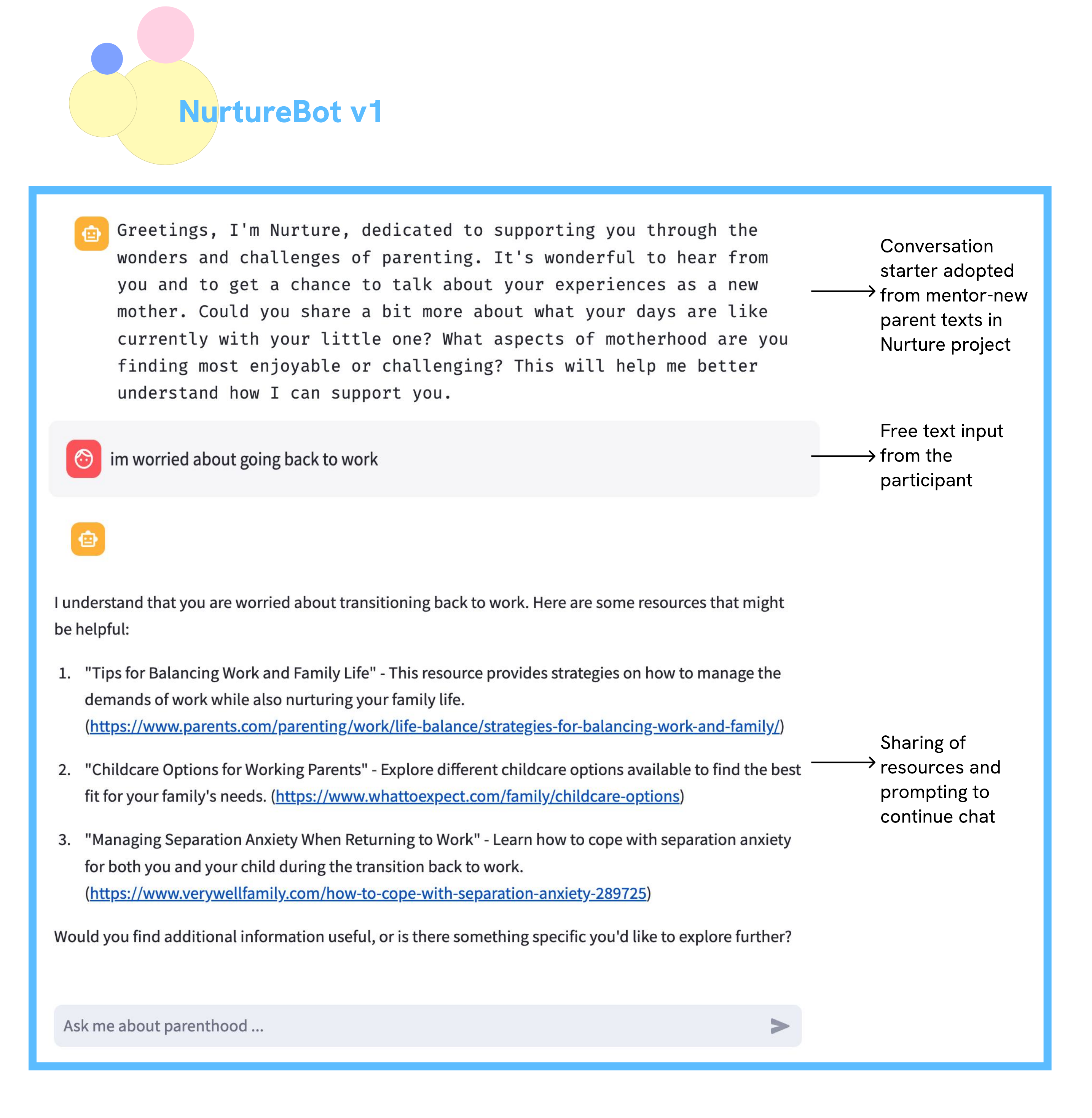}
  \caption{Example conversation with NurtureBot v1}
  \Description{xxx}
  \label{v1}
\end{figure}

Each day, participants interacted with the chatbot, receiving responses that included links to resources, empathetic messages, and suggestions for wellbeing exercises (see Figure \ref{v1}). At the end of each day, after interacting with NurtureBot hosted on a website via Streamlit \cite{streamlit}, participants were asked to complete a survey via Qualtrics \cite{qualtrics}, providing feedback on their experience with the chatbot by answering the questions listed in Figure \ref{survey}. On the final day, they also completed the Chatbot Usability Questionnaire (CUQ), a specialised questionnaire designed to access the usability of healthcare chatbots \cite{cuq}.

\subsubsection{Data Collection and Analysis}
\label{trialdataanalysis}
Over the course of the five-day study, 144 conversations and associated feedback were generated, denoting a daily-average of 90\% (28.8 out of 32 participants). All data was securely stored on our university's servers, compliant with GDPR \cite{EU2016GDPR}, and no personal data was collected, consistent with Prolific platform's design.

The qualitative analysis of the feedback from the daily surveys (See questions 1 and 2 in Figure \ref{survey}) was conducted using a thematic analysis approach \cite{braun2012thematic}. Our analysis focused on the instances that captured parent reactions to interacting with NurtureBot. The first author conducted the initial analysis using inductive, open coding, and regularly met with the second author to discuss emerging themes and explore alternative interpretations of participant responses. While participants did share positive feedback on several aspects of the chatbot, highlighting its usefulness and relevance, especially during early stages of parenthood, we do not explore the parents perspectives on the potential of LLM-based parental wellbeing support as this is reported in the literature \cite{ashraf2024systematic}, and to keep our focus on answering RQ1. Hence, we focused on analysing 131 of the 288 (45.48\%) pieces of feedback that highlighted negative interactions and suggestions for improvement.  This led us to the identification of 28 unique pain points, which were further grouped into seven overarching themes, converged in three subsequent meetings. 


\subsubsection{Findings of initial trial: Parent reactions to Nurture Bot v1}
\label{v1TA}

\textbf{Qualitative results}
Our seven resulting themes, listed below, highlighted the key problems participants faced in NurtureBot v1's design and interaction style, which closely mirror observations reported in recent literature on LLM-based wellbeing support tools \cite{seo2024chacha, kim2024mindfuldiary}. Given the similarity to established findings, we do not discuss these issues in detail here, and interested readers can find a more comprehensive account in Appendix A. Also, we further discuss a comparative analysis of our findings in the Results section (see Section \ref{results}).

\begin{itemize}
    \item \textbf{Problem 1: Unable to Chat} -- Participants found NurtureBot’s tendency to cut off conversation and redirect to external links disruptive.
    \item \textbf{Problem 2: Patronising and Robotic Response Tone} -- Responses felt overly scripted, offering platitudes rather than empathetic engagement.
    \item \textbf{Problem 3: Lack of Localised Resources} -- Suggested resources were not tailored to participants’ specific geographical contexts.
    \item \textbf{Problem 4: Memory and Continuity in Conversations} -- NurtureBot’s inability to register or recall user details contributed to impersonal interactions.
    \item \textbf{Problem 5: Generic Responses} -- Users received unhelpful, link-heavy replies that lacked specificity and personal relevance.
    \item \textbf{Problem 6: Transactional Nature of Conversations} -- Rapid, question-focused exchanges offered little opportunity for reflective or natural interaction.
    \item \textbf{Problem 7: Content Overload} -- Long, link-laden responses overwhelmed participants and hindered meaningful engagement.
\end{itemize}

\textbf{Quantitative results}
Our subjective questions regarding NurtureBot v1's task satisfaction received an average score of 4.29 out of 5, while interactive questions averaged 4.36 for understandability, 3.97 for controllability, and 3.88 for improvability. The standard questionnaire analysis resulted in a mean CUQ score of 85.4 out of 100, which, when compared to the CUQ benchmark of 68 \cite{cuq} in a one-sample t-test, was extremely statistically significant (t(29) = 7.11, p < 0.0001, mean difference = 17.40, 95\% CI [12.40, 22.40], SE = 2.45). These scores and their implications are discussed further in Section \ref{results}, alongside comparisons with NurtureBot v2 and v3.

Thus, leveraging insights from the Nurture project in Part I, we designed NurtureBot v1 and initiated iterative trials by testing the prototype with everyday users. After identifying seven key problems with parent-NurtureBot v1 interactions in Part I, our goal in Part II was to co-design by tasking users to prioritise problems and imagine ideal solutions.

\section{Part II: ARC Co-Design Study}
\label{part3}
To begin addressing RQ2 --\textit{ ``What key user needs must a parental wellbeing assistant address? How do parents describe their ideal vision of how the assistant should function, particularly regarding their understanding of its workings, control over it, and opportunities to improve the outcomes?,''}-- the second part of the study involved Asynchronous Remote Communities (ARC) co-design sessions \cite{ARC}, designed to directly engage participants in the iterative design process.  These sessions were facilitated using the Miro platform \cite{miro2011}, allowing the same 32 participants who tested NurtureBot v1 (see Subsection \ref{cohort1}) to return to collaboratively co-design solutions asynchronously, thus accommodating different schedules, total anonymity, and fostering broad participation. 

The ARC sessions involve two core activities over three sessions (1 week), prioritising, and addressing key issue(s) with NurtureBot v1. The average participation was 29.33 out of 32 (91.63\%) participants per day. To help familiarise participants with using Miro, we designed a training activity and instructional tips for using Miro on mobile devices, as well as providing direct messaging support via Prolific. 


\subsection{ARC Activity One: Diagnosing and Prioritising Pain Points}
\label{sec:ARC-act1}
The goal of the first activity was to establish which of the expressed difficulties were the biggest priority to address. Using the seven key problems from NurtureBot v1's technology trial (see Section \ref{v1TA}), we asked participants to prioritise these issues. By disclosing these key issues, the goal was to ensure that all participants could see the broader scope of the negative feedback gathered from others within their group. 

\subsubsection{ARC Activity One: Method}
\begin{figure}[t]
  \centering
  \includegraphics[width=\linewidth]{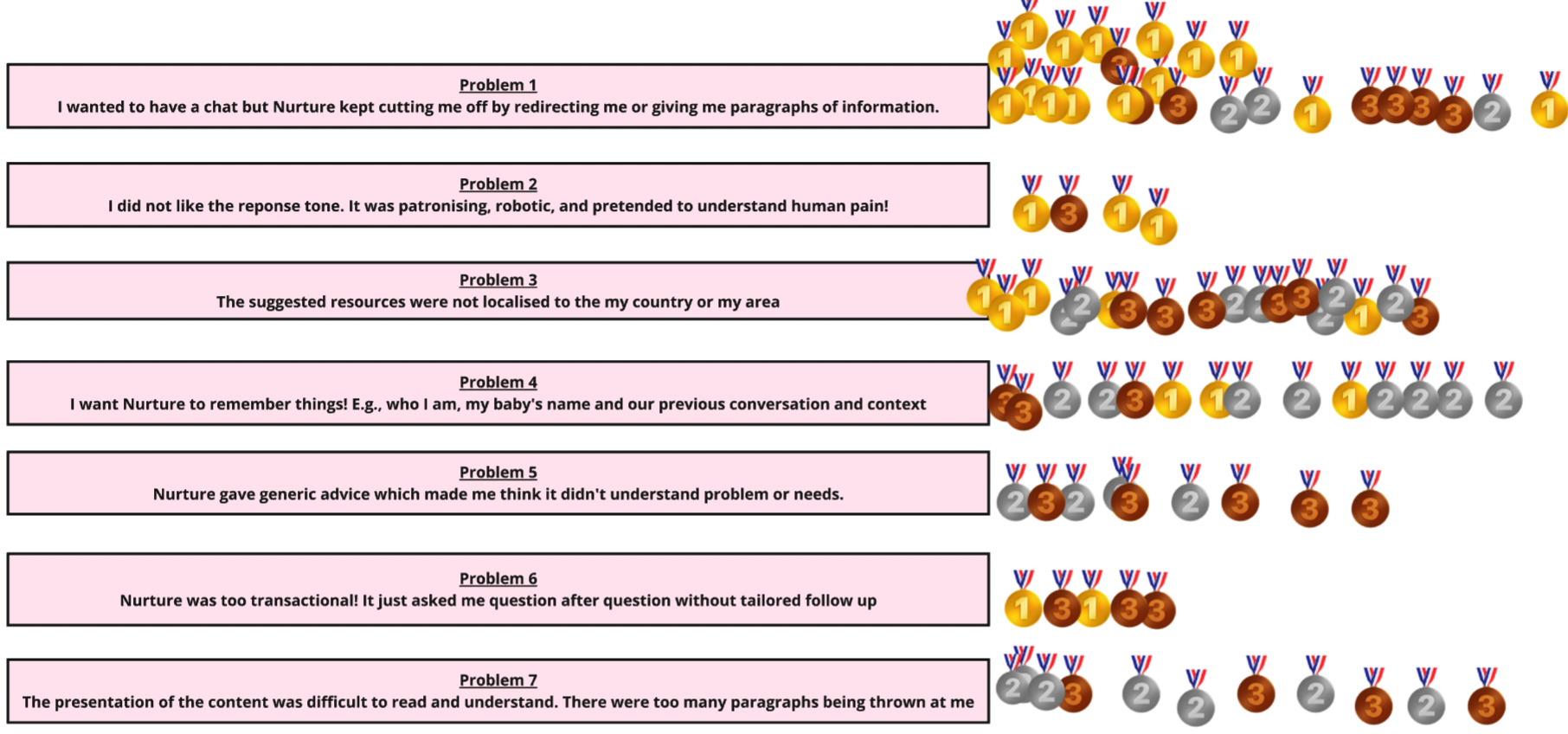}
  \caption{ Results from ARC Activity One voting by 30 participants, where key problems in interacting with NurtureBot were ranked using an Olympic points system. Complete Miro board including the other set-up phases participants went through before voting can be found in the supplementary materials.}
  \Description{xxx}
  \label{arcactivity1}
\end{figure}

We crafted a "problem statement" for each issue, synthesising the findings without directly replicating any participant's verbatim feedback (listed as problem statements in Figure \ref{arcactivity1}). This method not only helped verify our interpretation and synthesis of the feedback but also prevented potential bias, ensuring participants did not favour their own comments when voting. To provide sufficient scaffolding for reflection, we organised the activity as follows:

\textbf{Reviewing Identified Problems:} Participants were guided through a review of the seven major problems. As they reviewed these issues, they were asked to select one of three stickers—``This applies to me,'' ``This may apply to me,'' or ``This does not apply to me''—and place it on each of the seven problems. This step was important for setting a common understanding of the challenges identified from the group.

\textbf{Ranking and Prioritisation:} After reviewing the problems, as the next step, participants were asked to rank them according to their personal experiences, using stickers (gold, silver, and bronze icons provided on the Miro board) (See Figure \ref{arcactivity1}). This step was designed to capture participant perspectives on the perceived severity of each problem, which would inform the prioritisation of design efforts in subsequent activities.

\subsubsection{ARC Activity One: Results}
Based on participants responses, 
the following options were voted as the top three problems by participants:
\begin{itemize}
    \item \textbf{Problem 1 -} \textit{I wanted to have a chat but NurtureBot kept cutting me off by redirecting me or giving me paragraphs of information-} emerged as the top challenge with NurtureBot. 
    \item \textbf{Problem 3 -} \textit{The suggested resources were not localised to my country or area-} was the second priority. 
    \item \textbf{Problem 4 -} \textit{I want NurtureBot to remember things! E.g., who I am, my baby's name and our previous conversation and context-} was ranked third. 
\end{itemize}

We next took the top-ranked problem (Problem 1) to begin considering how we might enhance ``successful chat problem'' in Activity 2.

\subsection{ARC Activity Two: Co-Designing Solutions}
\label{sec:ARC-act2}
The second activity focused on solution generation for problem 1. 
Participants were asked to step into the role of NurtureBot: \textit{``Imagine You Are NurtureBot and Fill in the Blanks.''}



\subsubsection{ARC Activity Two: Method}
The activity was structured as follows:

\textbf{Scenario-Based Problem Solving:} As the first step, participants were presented with three specific scenarios corresponding to the three main features of NurtureBot (i) interacting with NurtureBot for the first time and wanting to have an empathetic chat, (ii) seeking wellbeing exercises to manage work stress, and (iii) seeking information to soothe their baby; that exemplified trying to have a successful chat aiming to use all three features of NurtureBot. 
For each scenario, participants reviewed an example response and were asked to consider how NurtureBot could help users to better understand its offerings, control the conversation, and improve its outcomes.

\begin{figure*}[h!]
    \centering
    \subfigure{
        \includegraphics[width=0.3\textwidth]{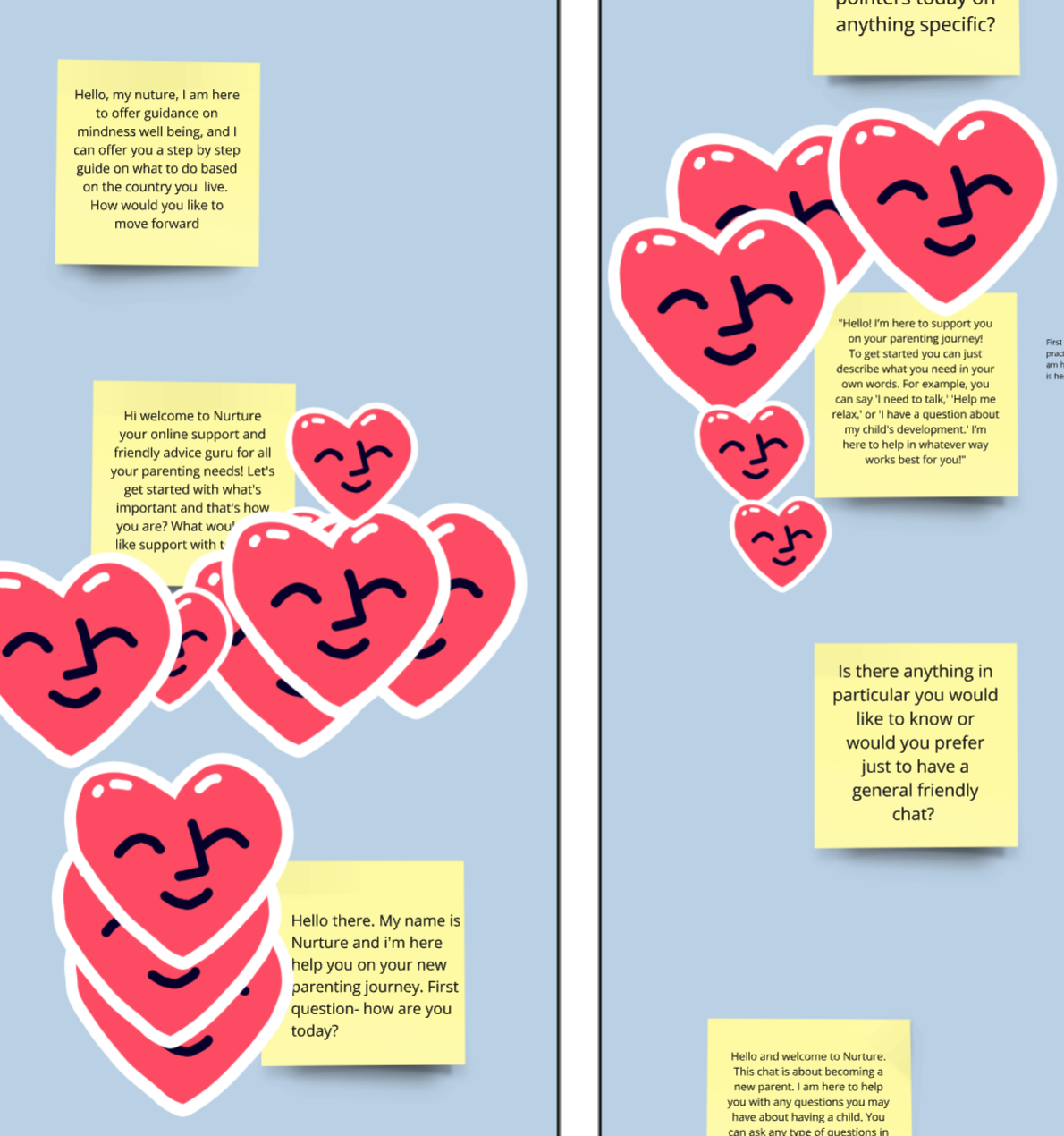}
    }
    \hfill
    \subfigure{
        \includegraphics[width=0.3\textwidth]{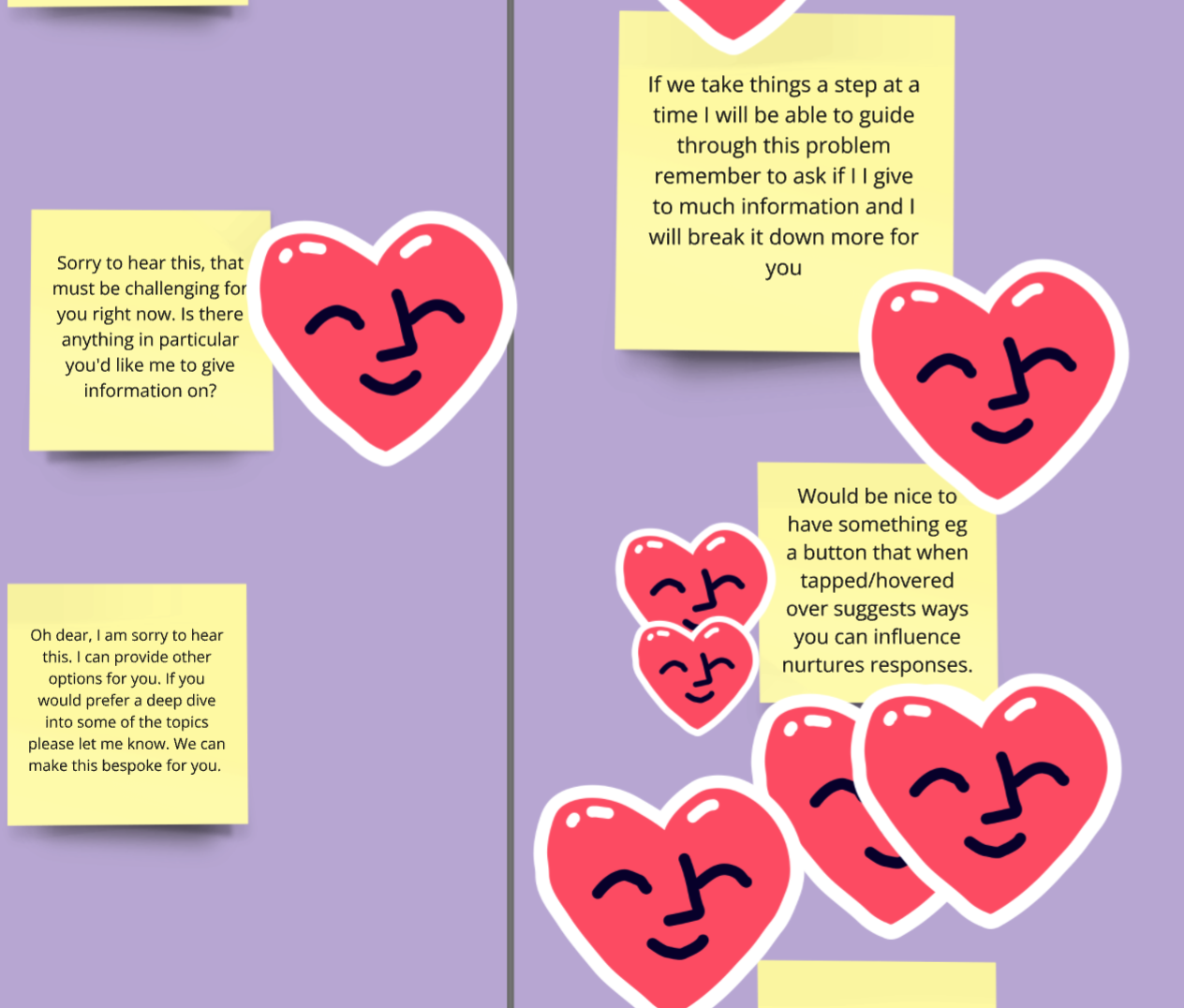}
    }
    \hfill
    \subfigure{
        \includegraphics[width=0.3\textwidth]{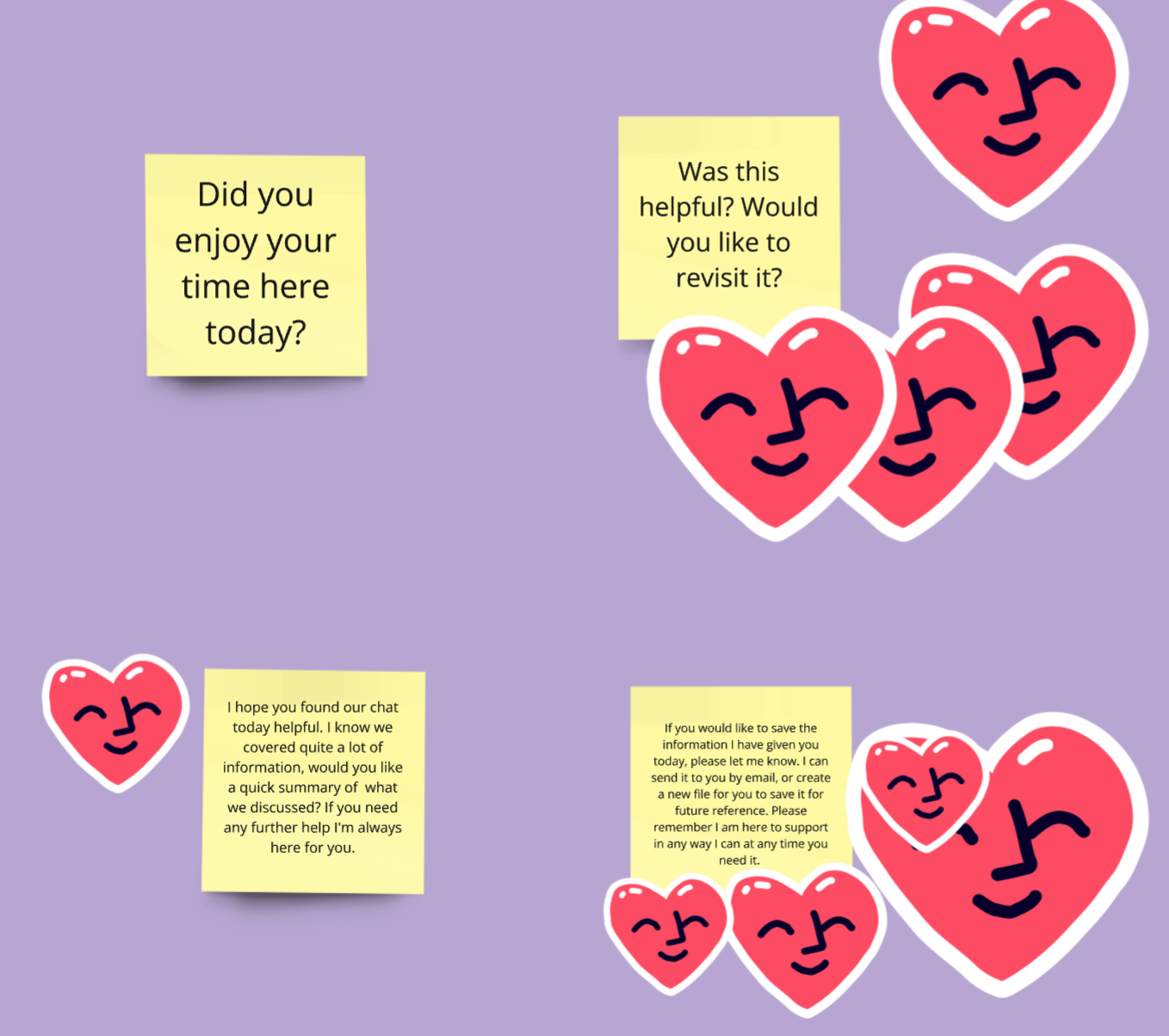}
    }
    \caption{A section from ARC Activity Two with 28 parents roleplaying as NurtureBot writing dialogues to better understand(left), control(centre), and improve(right) the features of empathetic chatting, wellbeing exercises, and parenting information lookup, and 30 parents returning to the Miro board to review and vote with heart stickers on the favourites among 189 dialogues generated. Complete Miro board including all dialogues and votes can be found in the supplementary materials.}
    \Description{xxx}
    \label{UCI}
\end{figure*}

\textbf{Collaborative Dialogue Writing:} Next, participants were invited to take on the role of NurtureBot and rewrite it's dialogue for each scenario. We offered cues that encouraged participants to work on "Understanding, Controlling, and Improving" NurtureBot, similar to the idea of a cybernetic loop \cite{patten1981cybernetic} (see Section \ref{HCAI}). This part of the activity encouraged participants to think critically about how the chatbot could better meet their needs in specific contexts, and which would inform NurtureBot v2 (See Figure \ref{UCI}).

\textbf{Iterative Feedback and Voting:} As the last step, after proposing solutions, participants reviewed and built upon each other's suggestions. We invited participants to vote on the best ideas (See heart stickers in Figure \ref{UCI}), ensuring that the most popular and potentially effective solutions were prioritised for implementation.


\subsubsection{ARC Activity Two: Results}

\sruthi{
ARC Activity Two generated a rich dataset of 189 participant-generated dialogues. 
Our thematic analysis of these dialogues (as in Section \ref{trialdataanalysis}) generated seven major user needs as follows:
\begin{itemize}
    \item \textbf{User Need 1: Use of metaphors for understanding} to make interactions more relatable and easier to comprehend.
    \item \textbf{User Need 2: Beginning with empathy} to establish trust and provide emotional support during interactions.
    \item \textbf{User Need 3: Introducing features clearly} to guide users effectively through the system's functionalities.
    \item \textbf{User Need 4: Providing step-by-step guidance} to ensure users can easily follow instructions or exercises.
    \item \textbf{User Need 5: Personalising and deepening sessions} to make the experience relevant and engaging for users.
    \item \textbf{User Need 6: Offering clarification and reassurance} to address user doubts and foster confidence in the system.
    \item \textbf{User Need 7: Summarising and offering follow-up options} to conclude conversations effectively and provide opportunities for further exploration.
\end{itemize}
We provide a comprehensive summary of these user needs in Appendix B and in Table \ref{tab:quotes_mapping}, we map examples from participant-generated dialogues onto the corresponding user needs.}

\begin{table*}[htbp]
\centering
\caption{Mapping Parent-Written Dialogues from ARC to NurtureBot's User Needs}
\label{tab:quotes_mapping}
\resizebox{\linewidth}{!}{%
\begin{tabular}{@{}p{0.48\linewidth} p{0.48\linewidth}@{}}
\toprule
\textbf{Dialogues written by parents} & \textbf{Mapped User Needs} \\ \midrule

“Hi welcome to NurtureBot your online support and friendly advice guru for all your parenting needs! Let's get started with what's important and that's how you are? What would you like support with today? :)” & 
User Need 1. Use of Metaphors for Understanding\newline
User Need 2. Beginning with Empathy \\ \midrule

“Remember, if you find that this exercise isn’t quite what you need, feel free to ask for an alternative. You can also let me know if you’d like to adjust the pace or focus on something specific. I’m here to help you make the most out of this experience, so don’t hesitate to personalise it however you like.” & 
User Need 2. Beginning with Empathy\newline
User Need 5. Personalising and Deepening Sessions\newline
User Need 6. Offering Clarification and Reassurance\newline
User Need 7. Summarising Exercises and Offering Follow-Up Options \\ \midrule

“Was this support helpful? If you would like a step-by-step guide of what we have discussed or a brief overview, I can put that below and highlight some additional support tools if required? Your happiness is important! I would love to support you further.” & 
User Need 4. Providing Step-by-Step Guidance\newline
User Need 6. Offering Clarification and Reassurance\newline
User Need 7. Summarising and Offering Follow-Up Options \\ \midrule

“Don't worry, you are not alone! It is very common to face this problem as a new parent. Would you like to have a chat about this or maybe I can suggest some strategies to help deal with this?” & 
User Need 2. Beginning with Empathy\newline
User Need 3. Introducing Features Clearly\newline
User Need 6. Offering Clarification and Reassurance \\ \midrule

“I hope you found our chat today helpful. I know we covered quite a lot of information, would you like a quick summary of what we discussed? You can save it to review later. If you need any further help, I'm always here for you.” & 
Feature 6. Offering Clarification and Reassurance \newline 
Feature 7. Summarising and Offering Follow-Up \\ \midrule

“I am here to help you through your journey as a new parent. It can often be overwhelming so if you have anything specific I can help with or you just need to use me as a sound-board—I am here, day or night.” & 
User Need 1. Use of Metaphors for Understanding\newline
User Need 2. Beginning with Empathy\newline
User Need 6. Offering Clarification and Reassurance \\ \midrule

“Hello! I’m here to support you on your parenting journey. To get started you can just describe what you need in your own words. For example, you can say ‘I need to talk,’ ‘Help me relax,’ or ‘I have a question about my child's development.’ I’m here to help.” & 
User Need 1. Use of Metaphors for Understanding\newline
User Need 2. Beginning with Empathy\newline
User Need 3. Introducing Features Clearly\newline
User Need 7. Summarising and Offering Follow-Up Options \\ \midrule

“I'm here to help answer or guide any questions you have about parenting, or how to manage the stress of it all. I'd like to know a bit more about you. How are you today?” & 
User Need 2. Beginning with Empathy\newline
User Need 3. Introducing Features Clearly\newline
User Need 5. Personalising and Deepening Sessions \\ \midrule

“You're in control and I will follow your lead so please let me know how you would like this conversation to go. (Would be nice to have a button to tap that suggests ways you can influence responses.)” & 
User Need 5. Personalising and Deepening Sessions\newline
User Need 7. Summarising and Offering Follow-Up Options \\ \midrule

“Parenting can be tough on the mind and body so let's explore ways as what gentle exercises can support us with. Would this be helpful for you?” & 
User Need 2. Beginning with Empathy\newline
User Need 3. Introducing Features Clearly\newline
User Need 6. Offering Clarification and Reassurance \\ \midrule

“If you need more clarifications on what I can do, please ask. I can present information in other formats such as tables and bullet points.” & 
User Need 4. Providing Step-by-Step Guidance\newline
User Need 6. Offering Clarification and Reassurance \\ \midrule

“There are a number of different tools that would help you find the right balance when facing these challenges. Would you like me to list all the different tools, or would you like to provide some more details on what specific kind of tools you are looking for?” & 
User Need 3. Introducing Features Clearly\newline
User Need 4. Providing Step-by-Step Guidance\newline
User Need 5. Personalising and Deepening Sessions \\ \bottomrule
\end{tabular}%
}
\end{table*}

Our findings from ARC in Part II, which elicited the key user needs required to have a successful chat with Nurture, were then directly employed to implement NurtureBot v2 in Part III.




\newpage
\section{Part III: The Interaction Layer: NurtureBot v2}
\label{part4}
Part III continued addressing RQ3: \textit{``How can we employ co-design methods to craft interactions around AI, enhancing usability and the overall user experience?.''} The development of NurtureBot v2 focused on refining interactions to enhance its user experience. \sruthi{Designing LLMs to respond empathetically and appropriately in sensitive domains such as parental wellbeing is a complex endeavour \cite{zamfrescu2023herding}, requiring meticulously crafted prompts and interaction architectures. Key challenges include effectively capturing user feedback, ensuring model adherence to instructions, and validating that system behaviour aligns with user needs.

To address these challenges, we developed \textit{the interaction layer} prompt architecture. This architecture builds on advanced prompting techniques, including Chain-of-Thought prompting \cite{wei2022chain}, which reveals intermediate reasoning steps, and ReAct \cite{yao2022react}, which integrates reasoning with action-based responses. These methods underscore the rapid evolution of prompt engineering towards interpretability, reliability, and alignment with user expectations. Our interaction layer operationalises these advancements, functioning as a mechanism to incorporate user feedback directly into the prompt while dynamically adapting system behaviour through state transitions. By conceptualising the interaction layer as a state machine, we linked human-centred design principles to actionable system behaviours, enabling the chatbot to flexibly adapt based on user input.}

We created an interaction layer central to how the LLM-based chatbot engages with users (see prompt illustrated in Figure \ref{prompt}). There were four aspects of NurtureBot v2's interaction layer, namely: 1) the \textbf{core interaction principle} of NurtureBot orchestrating the conversations through the understand, control, and improve model (as discussed below), 2) \textbf{interaction states} acting akin to a state machine (denoted by four big blue dots in Figure \ref{prompt}, 3) \textbf{three interaction levels} within each interaction state scaffolding instructions and examples for user understanding, controlling, and conversational improvement, and 4) \textbf{interaction elements} further embedded within interaction levels (denoted by multiple big yellow dots in Figure \ref{prompt}). All of these were directly shaped by the ARC co-design sessions and implemented using a structured prompt architecture that ensured consistency and depth in user interactions. Our design approach extends Beaudouin-Lafon’s theory of instrumental interaction \cite{MBL}, adapting it to LLM-based AI agents by creating four instruments within the interaction layer, invisible in the user interface, that reify user needs and are polymorphised, and reused to improve situational interaction across the prompt architecture.

\subsection{Interaction layer design}
\subsubsection{Core Interaction Principle}
NurtureBot v2's development was guided by three foundational capabilities we wanted users to have in relation to the chabtot, and vice-versa: \emph{understanding, control,} and \emph{improvement}. These core interaction principles (denoted with a yellow rectangle Figure \ref{prompt}) guided all interactions with users. This model, inspired by the concept of the `cybernetic loop' \cite{patten1981cybernetic}, and supported by the human-centred AI research endeavours discussed in Section \ref{HCAI}, was designed to ensure that every exchange was meaningful, user-centred, and adaptable to the individual and situational needs of each parent. This principle provided a template for each type of exchange, guiding the chatbot on how to initiate conversations, present options, and offer follow-up support.

\subsubsection{Three Interaction Levels}

\textbf{1. Understand Level:} The chatbot's first task was to understand and acknowledge the user's situation while helping the user grasp how the chatbot could assist them. For instance, it would ask, ``How are you feeling today?'' or ``Can I help you with a list of options?'' We employed the metaphors of ``AI support'' and ``AI assistant'' to minimise users anthropomorphising the chatbot, ensuring they maintained a clear understanding of its role as a helpful tool designed with empathy rather than attributing human relationships to it. This phase was essential for setting a supportive tone and ensuring the chatbot could convey its capabilities and tailor its responses to meet the user's needs effectively. 

\textbf{2. Control Level:} Once the user’s needs were understood, the chatbot entered the control level, where it provided options to guide the interaction along with its content. This level was designed to give users a sense of agency and control over the conversation. Users could choose to elaborate on their concerns, request a summary, or switch topics. Users would be often provided with "hints" such as revise, restart, and elaborate; designed in ARC sessions, in accordance with the context of the chat to enable them to effectively lead NurtureBot towards the desired outcome. 

\textbf{3. Improve Level:} After addressing the user’s needs, NurtureBot gave a summary and sought feedback to access user satisfactions and improve the interaction by offering additional resources or to revise the conversation (as parents preferred). This phase ensured that the conversation concluded on a constructive note, always asking for user feedback, with the user feeling supported and informed.

\begin{figure*}[!htbp]

  \centering
  \includegraphics[width=0.95\textwidth,height=0.95\textheight]{img/prompt.pdf}
  \caption{This figure illustrates the interaction layer prompt architecture of NurtureBot v2, integrating user-generated dialogues from ARC as few-shot examples in the prompt.}
  \Description{A detailed diagram of NurtureBot v2’s prompt architecture hosting an interaction layer, divided into core principles: Understand, Control, and Improve. It highlights the chatbot’s empathetic chatting, wellbeing exercises, and parenting information, showing how the system tailors responses based on user preferences and inputs. Examples of user dialogues, actionable hints, and resources are shown alongside elements for adapting to contextual user needs, illustrating NurtureBot’s design for personalised support.}
  \label{prompt}

\end{figure*}

\subsubsection{Interaction States and Elements}
At any time, NurtureBot could be in one of four interaction states, featuring specific interaction elements within each:

\textbf{1. Undecided Exchanges} This state was the default state which enabled NurtureBot to elicit from users what they want, and redirect NurtureBot into another state. When users were undecided, NurtureBot aimed to first understand their hesitation or indecision. It started with elements such as, ``What’s on your mind today?'' or ``How can I assist you?'' to encourage the user to reflect on their needs.
Metaphorical elements like ``assistant'' were also embedded in the chatbot's dialogue to offer a supportive role, and reassurance elements were designed to not overwhelm user with options, reflecting a tone of gentle guidance. For instance, ``I’m here to support you in whatever way works best for you'' to help users feel comfortable in sharing their thoughts or challenges.

\textbf{2. Empathetic Chatting:} The chatbot was programmed with elements to focus on active listening and providing emotional support. We used dialogue examples to create non-judgemental elements and avoiding any advice. The chatbot’s responses were crafted to mirror those of a trusted confidant, providing reassurance and understanding.

For example, these elements instructed NurtureBot to initially wait: ``Tell me more about what's been on your mind today.'' Based on the ARC study insights, the chatbot would then offer subtle prompts to encourage to-and-fro discussion, e.g., ``That sounds really challenging. How has that been affecting you?''.

\textbf{3. Wellbeing Exercises:} In this state NurtureBot provided guiding elements for wellbeing exercises included mindfulness practices and relaxation techniques. The interaction was structured with  personalised elements, offering tailored exercises based on user concerns and alternative options. The chatbot also ensured that exercises were broken down into manageable steps, with the user controlling the pace of the session. To embed collaborative elements, the prompts also ensured that NurtureBot could guide users through activities step by step, e.g., ``Would you like to try a simple breathing exercise together? Let me know when you're ready to begin.'' As such, the user could control and personalise the pace and depth of the exercise.

\textbf{4. Parenting Information:} NurtureBot now provided a structured and digestible format for parenting content. Instead of overwhelming users with lengthy text, interactive elements offered summaries and gave users the option to dive deeper into specific topics. The chatbot also cross-referenced reliable sources to ensure that the information was accurate and relevant. The offered information was also adapted, based on the specific situation, preference, and needs. For instance, if a user asked about managing a fussy baby, the chatbot might respond, ``I have a few techniques that could help. Would you like a quick overview, or shall we talk about your baby and things you have tried already?'' This ensured that users received information that was most useful to them.

It is to be noted that that aside from the changes made to the prompt architecture as described above, no additional modifications were introduced in version 2 in comparison with version 1 (see Figure \ref{promptarch}). This was intentional, as the conversational dataset collected during the testing of NurtureBot v1 was fraught with issues discussed in Section 4.3.1. At this stage of development, we chose to focus on testing the efficacy of our human-centred prompt architecture, the interaction layer, without training/tuning of the model.



\subsection{Technology trial of NurtureBot v2}
\label{part5}

\subsubsection{Participants, Method, and Data}
The same initial set of participants (See Section \ref{cohort1}) trialled NurtureBot v2 over five days, generating a total of 104 conversations and their corresponding survey feedback (See Figure \ref{survey}). We analysed the survey data in accordance with our thematic approach (See Section \ref{trialdataanalysis}). Participation averaged 69.3\% (20.8 out of 30), a decline from 90\% in Part I (v1 trial) and 91.6\% in Part II (ARC session). Eight participants contacted via Prolific cited reasons such as, ``I've enjoyed your study, but NurtureBot has improved so much I don't know what more I can give feedback on.''

\subsubsection{Findings of second technology trial - NurtureBot v2}

\paragraph{Qualitative results}\quad While our data revealed highly positive feedback for NurtureBot v2, with participants praising the chatbot for being ``more chatty'' and ``engaging'' than its predecessor, as with NurtureBot v1, we only focused the key interaction problems for our thematic analysis, focusing on 78 instances of negative feedback and/or suggestions for improvement (of 208 instances of feedback collected (37.5\%)), coding 12 unique pain points, which we then grouped into three themes briefly summarised below. We further discuss a comparative analysis of our findings in the Results section (see Section \ref{results}).


\textit{Problem 1:} Some participants pointed out that NurtureBot still occasionally responded with long paragraphs, making the information feel overwhelming. One participant noted, \begin{quote} ``It was good, but very long answers though and lots of information to take in.'' \end{quote}

\textit{Problem 2:} Most participants noted a bug, where it repeatedly suggested the same prompts to the user—``Do you wish to elaborate, revise, or restart?''—which reduced the creativity and variety in responses. A participant mentioned, \begin{quote} ``The repetition of the last sentence `if you need me to elaborate... just let me know' was quite annoying.'' \end{quote}

\textit{Repeating Themes:} As with v1, participants made suggestions for incorporating buttons for conversational hints, multi-modal content (images and videos), memory capabilities for further personalisation, and connection to resources and communities in their local area.

\paragraph{Quantitative results.}\quad Our subjective questions regarding NurtureBot v2’s task satisfaction received an average score of 4.41 out of 5, while interactive questions averaged 4.50 for understandability, 4.42 for controllability, and 4.44 for improvability. The questionnaire analysis resulted in a mean CUQ score of 87.2 out of 100, which, when compared to the CUQ benchmark of 68 \cite{cuq} in a one-sample t-test, was extremely statistically significantly (t(20) = 8.98, p < 0.0001, mean difference = 19.20, 95\% CI [14.74, 23.66], SE = 2.14). These scores and their implications are discussed further in Section \ref{results}, alongside comparisons with NurtureBot v1 and v3.

While our findings denoted raised user satisfaction, it is to be noted that only participants who were open to the concept of a chatbot solution responded positively to NurtureBot. Among the 32 participants in the first cohort, one participant (approximately 3\%) continued to comment that they would prefer using search engines over interacting with NurtureBot. This participant did not like the chatbot solution from the beginning and remained unchanged in their preference.

In sum, NurtureBot v2, featuring an interaction layer prompt architecture created in Part III with parent-designed dialogues, and was tested to ensure that the issues identified in the previous version were effectively addressed. The two minor interaction issues persisting in NurtureBot v2, identified in Part III, were targeted for refinement in NurtureBot v3 and tested with a new cohort of participants in Part IV to further validate our efforts.

\section{Part IV: Prototype Improvement: NurtureBot v3 and Validation}
\label{part6}

\subsection{The Final Version: NurtureBotv3} 
\label{v3}

Based on the feedback from Part III, we iterated a final version, NurtureBot v3 by making three minor changes in the interaction layer as briefly summarised here and illustrated in comparison with the prompting techniques of NurtureBot v1 and v2 in Figure \ref{promptarch}:
\begin{itemize}
    \item First, we addressed the issue of lengthy responses by limiting the chatbot's answers to 50 words, ensuring information was delivered in concise, digestible segments. 
    \item Second, we updated the core interaction principle, enhancing NurtureBot’s ability to provide contextual actionable hints, rather than offering generic options (denoted as struck-out text inside the big yellow rectangle in Figure \ref{prompt})
    \item Third, as the last step of the prompt we included a pre-output validation to assess formatting and contextual relevance of the hints suggested to the user, ensuring LLM's adherence to the first two changes.
\end{itemize}

\sruthi{It is to be noted that apart from these three minor changes made to the prompt architecture as described above, no further modifications were implemented in NurtureBot v3 compared to version 2 (see Figure \ref{promptarch}). This decision was intentional; despite having a robust conversational dataset from the NurtureBot v2 testing phase, our primary focus was to evaluate the effectiveness of the interaction layer—with a new sample of participants. This approach allowed us to isolate and validate the impact of the co-designed prompt architecture.}

\sruthi{
\paragraph{Final version overview:}\quad At the end of our iterative design process, the final version, NurtureBot v3, incorporated a co-designed and parent-tested interaction layer. This human-centred prompt architecture empowered the chatbot to sustain reliable, context-aware conversations. By leveraging parent-generated few-shot examples from the ARC co-design sessions, NurtureBot could dynamically generate dialogues tailored to diverse user needs. The seven user needs identified at the end of ARC co-design sessions \ref{sec:ARC-act2}, were systematically mapped onto the interaction layer prompt architecture, enabling the chatbot to generate dialogues styles that matched user expectations (see Figure \ref{convo}), ensuring a user-aligned interaction experience.

As illustrated in Figure \ref{convo}, the conversation between the new parent and NutureBot, always orchestrated by the core interaction principle, begins with an explanation of its capabilities alongside an wellbeing check-in, setting a supportive tone and helping parents understand how the system could assist them. 
Based on the parent's response, the interaction then moves into one of the four interaction states. For example when the parent expresses a positive or negative wellbeing response (as in Figure \ref{convo}) it moves to the undecided exchange state, where further interaction levels of Understand, Control, Improve come into play in one or more exchanges, and NurtureBot invokes interaction elements specially designed for this state by offering situational reassurance and reiterating its core functions to enhance user understanding. It also provides tailored hints to empower the parent with actionable phrases to drive the conversation towards meaningful engagement while offering clarifications and follow-ups as needed to further improve the conversation. When the user selects a specific feature—for example, parenting information (as in Figure \ref{convo})—the interaction state transitions accordingly. Further into this state, the interaction follows the same architecture again via the three interaction levels and many state-specific interaction elements. NurtureBot validates its understanding of the parent’s context, and aims for personalising and analysing the information provided to ensure relevance and precision. Following this, it offers structured resources, providing options to explore the information further or save it for future reference. The conversation concludes with an explicit request for feedback, which reinforces the chatbot’s adaptability and commitment to continuous improvement. This layered approach highlights the adaptability of the system's interaction model, the interaction layer, which governed by the core interaction principle, dynamically adjusts states, levels, and elements based on user inputs, ensuring a seamless and supportive conversational flow.

}
\begin{figure*}[!htbp]

  \centering
  \includegraphics[width=0.95\textwidth,height=0.95\textheight]{img/Conversation.pdf}
  \caption{Example of a conversation with NurtureBot v3 featuring a co-designed and iterated interaction layer prompt architecture.}
  \Description{}
  \label{convo}

\end{figure*}

\subsubsection{Participants, Method, and Data} 
To assess the effectiveness of these improvements without learning and co-design biases, NurtureBot v3 was tested with a new cohort of 46 parents, 29 females and 17 males with an average age of 38.07. 11 participants were already familiar with chatbots, with (23.9\%) using ChatGPT daily. A total of 185 conversations (denoting an average participation of 80.4\% (37 out of 46 parents across all days)) and corresponding feedback were collected over five days using the same data collection (See Figure \ref{survey}) and analysis method (see Section \ref{trialdataanalysis}) as outlined in the NurtureBot v1's trial, allowing for direct comparisons across versions.

\subsubsection{Findings of third technology trial - Nurture Bot v3}
Next we present a summary of the third trial findings, focusing on identifying ongoing interaction problems, following the interaction layer redesign.

\paragraph{Qualitative Results}\quad While feedback was overwhelmingly positive, our analysis focused the interaction problems (from 46 out of 370 (12.4\%) negative feedback. We identified 10 unique pain points, which we group into the three themes, summarised below. We also provide a comparative analysis of the three Nurturebot versions in Section \ref{results}.

\textit{Problem 1:} Some participants wanted NurtureBot to ``intelligently'' provide more detailed responses when necessary. For example, \begin{quote} ``I guess the responses could be a little longer and detailed at times but not by much.'' \end{quote}

\textit{Problem 2:} Few participants found it challenging to follow step-by-step exercises via text. One participant remarked, \begin{quote} ``A lot of the time the advice was to- close my eyes then do X, but when it's text based this is obviously hard to follow. Perhaps tell them to read the advice then close eyes.'' \end{quote}

\textit{Repeating Themes:} Participants continued to request features for: multi-modal interactions (video, speech, and images), improved memory function to allow NurtureBot to remember personal details and past conversations, local resources, and hints to continue conversation as buttons. Furthermore, one participant expressed interest in viewing the original sources referenced in NurtureBot's responses, allowing them to verify or explore the information further.


\paragraph{Quantitative results.}\quad Our subjective questions regarding NurtureBot v3’s task satisfaction received an average score of 4.49 out of 5, while interactive questions averaged 4.63 for understandability, 4.04 for controllability, and 4.21 for improvability. The questionnaire analysis resulted in a mean CUQ score of 91.3 out of 100, which, when compared to the CUQ benchmark of 68 \cite{cuq} in a one-sample t-test, was extremely statistically significant (t(35) = 18.39, p < 0.0001, mean difference = 23.30, 95\% CI [20.73, 25.87], SE = 1.27). These score and their implications are discussed further in Section \ref{results}, alongside comparisons with NurtureBot v1 and v2.




\begin{table*}[t]
\centering
\caption{Comparison of User Experience with Subjective Questionnaire across NurtureBot's Versions}
\label{tab:results}
\resizebox{\textwidth}{!}{%
\begin{tabular}{|l|l|l|l|}
\hline
\textbf{Question} & \textbf{GEE Model} & \textbf{Kruskal-Wallis} & \textbf{Dunn's Test (p-values)} \\
\hline
1. Meeting hopes and goals & 
V2: $\beta$=0.1257, p=0.284 & 
H=5.8539 & 
\multirow{2}{*}{Not performed} \\
& V3: $\beta$=0.2015, p=0.061 & p=0.0536 & \\
\hline
2. Understanding functionality & 
V2: $\beta$=0.1377, p=0.146 & 
H=11.1215 & 
V1-V2: p=0.727 \\
& V3: $\beta$=0.2721, p=0.001** & p=0.0038** & V1-V3: p=0.003** \\
& & & V2-V3: p=0.225 \\
\hline
3. Control over actions & 
V2: $\beta$=0.4521, p<0.001** & 
H=13.5127 & 
V1-V2: p<0.001** \\
& V3: $\beta$=0.0666, p=0.573 & p=0.0012** & V1-V3: p=0.757 \\
& & & V2-V3: p=0.017* \\
\hline
4. Improving performance & 
V2: $\beta$=0.5582, p<0.001** & 
H=19.9734 & 
V1-V2: p<0.001** \\
& V3: $\beta$=0.3256, p=0.005** & p<0.0001** & V1-V3: p=0.007** \\
& & & V2-V3: p=0.201 \\
\hline
\end{tabular}%
}
\caption*{Note: * p < 0.05, ** p < 0.01. V1, V2, and V3 represent Version 1, Version 2, and Version 3 of NurtureBot, respectively. GEE coefficients ($\beta$) represent the change relative to Version 1.}
\label{tab:quant}
\end{table*}

\subsection{Results: NurtureBot v1 vs. NurtureBot v2 vs. NurtureBot v3}
\label{results}

\subsubsection{Qualitative Results}
The comparison of qualitative results from NurtureBot v1, v2, and v3 reveal a clear progression in user satisfaction across versions, supported by a reduction in negative feedback gathered from our analysis, 45.48\% in NurtureBot v1 to 12.4\% in NurtureBot v3. While our participants praised the ``natural flow'' of conversations and the ``supportive, empathetic tone'' of our improved versions, repeating themes for features like multi-modal interactions and enhanced memory for personalisation and localisation persisted across versions. In this section we detail our qualitative findings from evaluations of NurtureBot's evolutions.


\textbf{A Successful Chat}
In NurtureBot v1, participants found the chatbot useful but expressed frustration with its tendency to cut off conversations, leading to a lack of conversational flow. Several participants noted that the responses were often patronising or generic. Informed by user feedback and ARC co-design, NurtureBot v2 introduced an \textit{interaction layer} centred on the principles of \textit{Understand}, \textit{Control}, and \textit{Improve}, enabling the chatbot to better grasp user needs, offer greater conversational control, and dynamically adapt across all features, giving users more autonomy and flexibility. We highlight these adapted elements in bold.

Participants expressed satisfaction with NurtureBot v2, noting improvements in how \textbf{natural and intuitive} the conversations felt. One participant shared, ``Good, lots of ideas for activities for playtime for my baby which was good, and Nurture continued to ask if I wanted further information at each step.'' Participants also liked the system's \textbf{flexibility}, saying, ``The interaction was easy again, it feels like Nurture can handle whatever you ask it, whether it's just chatting or asking for specific information.'' Participant feedback also emphasised how the interaction layer made the \textbf{conversation flow} seamlessly, with one participant noting, ``Interaction was the best this week—the responses from Nurture seemed natural and clear.'' The introduction of the interaction layer also improved the \textbf{clarity and engagement} of the system. One participant mentioned, ``It was really good. I found the introduction really helpful, particularly the clarification of the three kinds of things that Nurture could do.'' Another shared, ``Great! It made such a difference having the prompt to tell Nurture what I wanted.'' This interaction layer enabled participants to feel \textbf{more in control} and allowed the chatbot to adapt to user input, creating a \textbf{personalised}, engaging experience. 

Although NurtureBot v2 was tested by participants who had co-designed it, NurtureBot v3 was tested with a new cohort, which confirmed the success of the interaction layer. One participant remarked, ``Good. Conversation flows well, and everything provided was useful,'' while another stated, ``So seamless yet educative.'' Our findings suggest interaction layer to played a crucial role in ensuring that NurtureBot responded dynamically to user input, leading to an improved overall experience.

\textbf{The Time Effect}
A key observation from the five-day testing period was the gradual shift in participants' attitudes toward NurtureBot. Initially, participants were hesitant and skeptical, viewing it as an unfamiliar tool and doubting its ability to provide meaningful support. However, as they continued using NurtureBot over consecutive days, their perceptions became more positive, with many describing it in personal terms—some even referring to it as a ``friend always there'' to support them.
For instance, one participant testing NurtureBot v1 described their early experience as “detached and robot-like,” but by day five, they said, ``After using Nurture quite a few times now, it just feels like I'm chatting with someone that feels familiar and I have already built up a bit of a relationship with. It doesn't feel artificial,'' even when we had not changed anything in our implementation. The time effect led many participants to believe that NurtureBot was being updated continuously each day. One participant testing NurtureBot v3 commented, ``Wow, it was so much better today. I wrote quite a lengthy update, and it acknowledged every part, even pinpointing what was bothering me.'' Participants began using personal pronouns like "they" to describe NurtureBot over time. On day four, one participant remarked, ``I think it is getting there with becoming more personal, maybe ask what the user’s name is, so they could talk using your name,'' reflecting an increased emotional connection with the chatbot. This progression illustrates a time effect, where repeated interactions led to increased comfort, attachment, and even anthropomorphisation of the technology. The more time participants spent engaging with NurtureBot, the more they saw it as a supportive companion rather than just a tool. 



\textbf{Community and Peer Support}
Tying back to the origins of the Nurture project, which was a community-based peer-to-peer mentoring program by Committee for Children \cite{CommitteeForChildren} (as described in Section \ref{part1}), it is clear that the concept of community and peer support remains a significant need for participants using NurtureBot. Throughout the three versions of NurtureBot, participants frequently searched for local activities, community support centres, and peer forums. This theme of seeking peer support and localised resources also emerged as the second biggest problem identified in the ARC (see Section \ref{arcactivity1}). Additionally, several participants expressed a desire for more localised community resources, indicating the importance of having access to peer support networks within their geographical area. As one participant trying NurtureBot v2 remarked, ``When I asked for classes in my local area it could not help, that would be a good feature.'' and another trying NurtureBot v3 said ``I realised when asking a question about services in my local area that it does not know my local area.'' This feedback underscored the value of community-driven support structures.



\textbf{Personalisation and Contextualisation}
The third biggest challenge prioritised in our ARC session \ref{sec:ARC-act1}, the lack of personalisation and contextualisation in the chatbot's responses, also emerged throughout the testing of all three versions. Participants repeatedly expressed the desire for NurtureBot to remember personal details such as their name, their child's name, their unique circumstances, and previous conversations. Participant feedback such as, “Remembering my problems would still be nice,” and “The bot could ask our names and baby's details to help make it more personal,” in NurtureBot v2 and NurtureBot v3 echoed this need, emphasising how critical personalisation was to creating a more meaningful user experience.

\subsubsection{Quantitative results} 
\begin{figure}[t]
  \centering
  \includegraphics[width=250px]{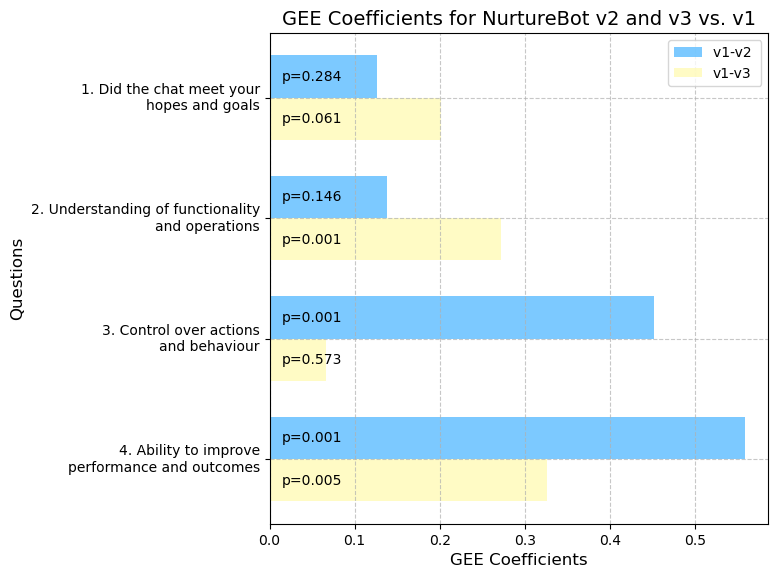}
  \caption{Generalised Estimating Equation (GEE) coefficients comparing chatbot versions v2 and v3 to v1 of NurtureBot across four evaluation dimensions: (1) whether the task met user expectations, and interactive dimensions of: user- (2) understanding, (3) control, and (4) ability to improve performance and outcomes. Statistically significant improvements (p < 0.05) are noted in dimensions (2), (3), and (4) in NurtureBot v2 and v3.}
  \Description{}
  \label{gee}
\end{figure}

\begin{figure}[t]
  \centering
  \includegraphics[width=250px]{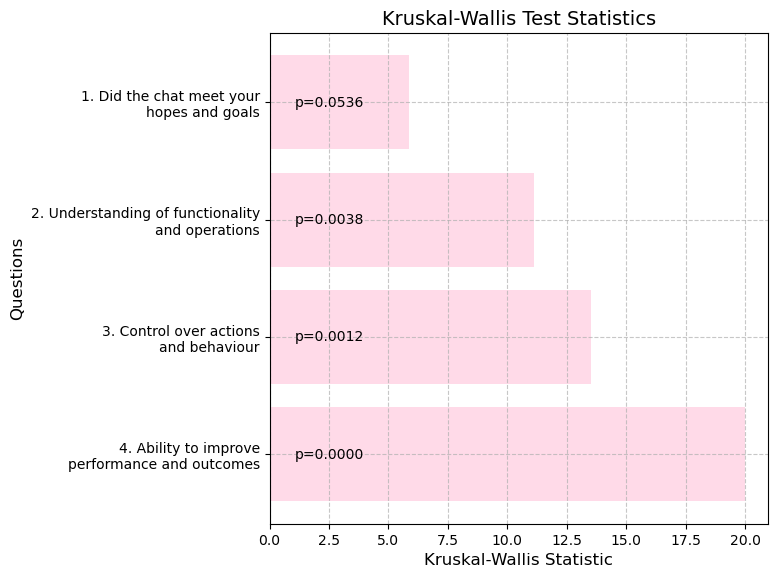}
  \caption{Kruskal-Wallis test results comparing the NurtureBot's versions across four dimensions of subjective user experience. The test identifies statistically significant differences (p < 0.05), with the strongest significance observed in the "ability to improve performance and outcomes" dimension.}
  \Description{}
  \label{kw}
\end{figure}

\begin{figure}[t]
  \centering
  \includegraphics[width=\linewidth]{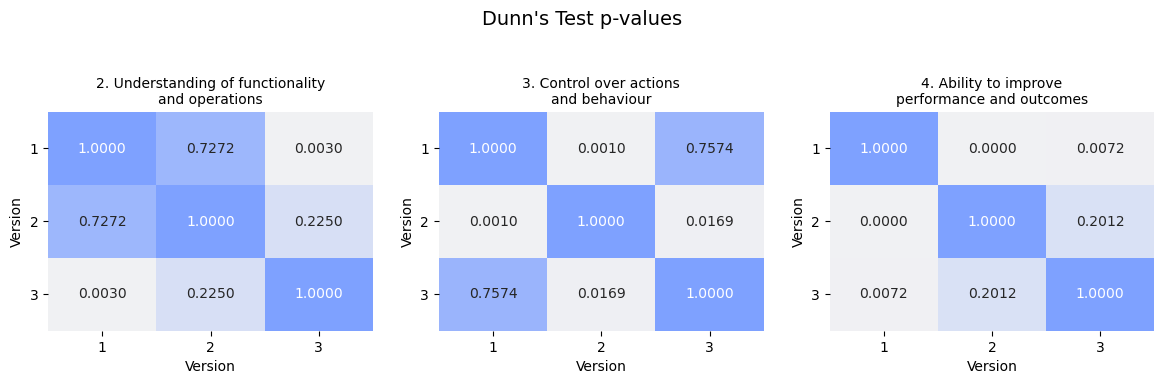}
  \caption{Heatmaps illustrating the pairwise comparison of three versions of the chatbot across three evaluation dimensions: (user- (2) understanding, (3) control  and , and (4) ability to improve performance and outcomes. Statistically significant differences (p < 0.05) are observed primarily between version 1 and version 3 across the evaluated dimensions. A reduced variance as NurtureBot is iteratively developed, also indicates chatbot's reliable performance over time.}
  \Description{}
  \label{dunns}
\end{figure}

To analyse our \textbf{subjective questionnaire} with four questions we created to measure task satisfaction and human-AI interaction (see questions 3-6 in Figure \ref{survey}). To the data we collected for these questions, across the three technological trials (in Sections \ref{part2}, \ref{part5}, and \ref{v3}), we employed both parametric Generalised Estimated Equations (GEE) and non-parametric Kruskal-Wallis and Dunn's test approaches (see Table \ref{tab:quant} and Figures \ref{gee}, \ref{kw}, and \ref{dunns}). Our results suggest:

\begin{enumerate}
    \item \textbf{Meeting User Expectations:}
    GEE showed no significant differences between versions (p > 0.05). Kruskal-Wallis yielded marginally significant results (H = 5.8539, p = 0.0536), suggesting potential subtle differences not captured by the parametric model. However, the changes between v1, v2, and v3 are not robust enough to be considered significant.

    \item \textbf{Understanding Functionality:}
    Both parametric and non-parametric tests indicated significant improvements. GEE showed Version 3 to significantly outperformed Version 1 ($\beta$ = 0.2721, p = 0.001). Kruskal-Wallis and Dunn's test confirmed this finding, revealing a significant difference between Version 1 and Version 3 (p = 0.003). 

    \item \textbf{Control Over Actions:}
    GEE indicated that Version 2 significantly improved user control compared to Version 1 ($\beta$ = 0.4521, p < 0.001), which was evident considering v2 was designed by v1 participants. Interestingly, Kruskal-Wallis and Dunn's test not only confirmed this (p < 0.001) but also revealed a significant difference between Version 2 and Version 3 (p = 0.017) that wasn't apparent with GEE.

    \item \textbf{Improving Performance:}
    Both analytical approaches showed significant improvements in later versions. GEE model indicated that both Version 2 ($\beta$ = 0.5582, p < 0.001) and Version 3 ($\beta$ = 0.3256, p = 0.005) significantly outperformed Version 1. Kruskal-Wallis and Dunn's test corroborated these findings, showing significant differences between Version 1 and both Version 2 (p < 0.001) and Version 3 (p = 0.007).
\end{enumerate}

 While the parametric test provided insights into overall trends, the non-parametric tests revealed additional pairwise differences, particularly between Version 2 and Version 3 in terms of user control, where we had changed the bug of repeating conversational hints. This comprehensive analysis suggests that the iterative development of NurtureBot has led to significant improvements in user understanding and ability to improve performance, with some complex dynamics in user control across versions.

\begin{figure}[t]
  \centering
  \includegraphics[width=250px]{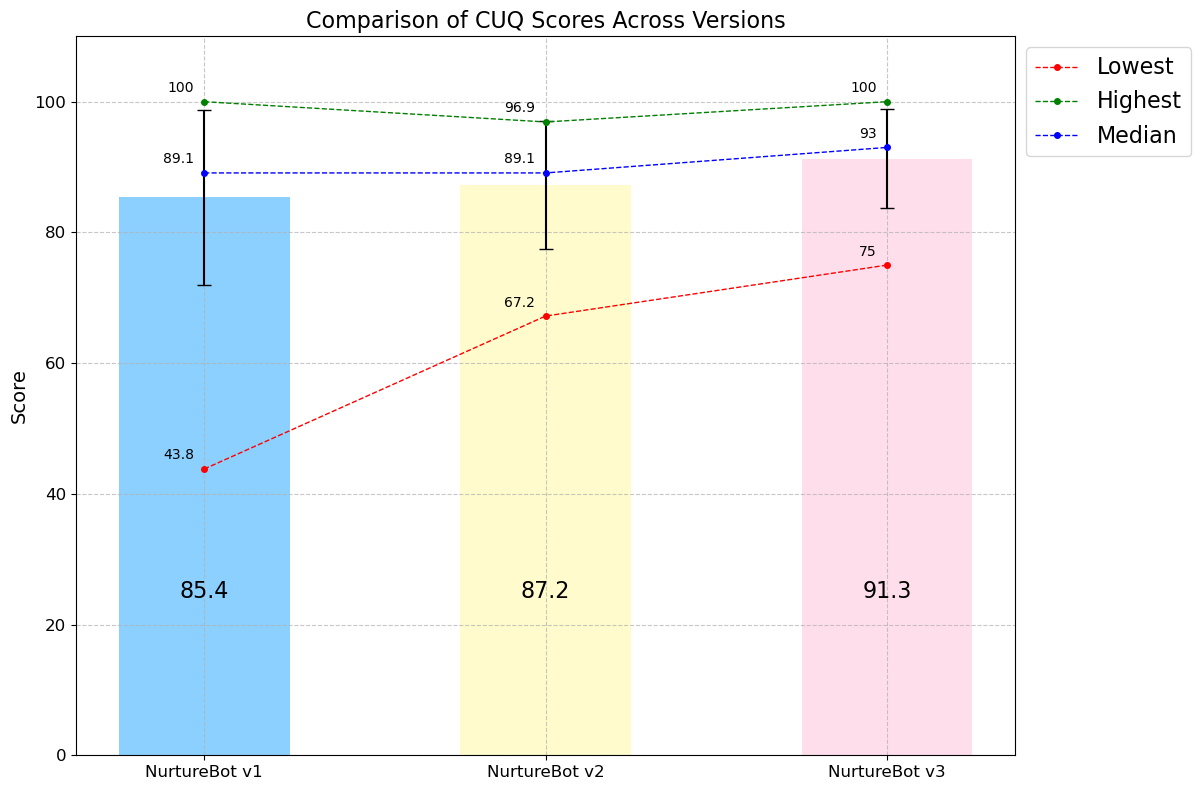}
  \caption{Comparison of CUQ Scores Across NurtureBot Versions. The figure demonstrates the improvement in Chatbot Usability Questionnaire (CUQ) scores from NurtureBot v1 with a CUQ score of 85.4, to NurtureBot v2 with a CUQ score of 87.2, and  NurtureBot v3, which achieved a higher CUQ score of 91.3. The lowest scores also improved significantly across versions, indicating enhanced chatbot usability with each iteration. }
  \Description{A graph comparing the CUQ scores of NurtureBot v1, v2, and v3. The x-axis lists the chatbot versions, while the y-axis displays the CUQ scores from 0 to 100. NurtureBot v1 has a score of 85.4, v2 scored 87.2, and v3 achieved the highest score of 91.3. The graph shows the lowest, median, and highest scores for each version, with a steady increase in the lowest scores from v1 (43.8) to v3 (75), while the median and highest scores remain consistently high, demonstrating improvements in chatbot usability over time.}
  \label{cuq}
\end{figure}

\begin{figure*}[!htbp]
  \centering
  \includegraphics[width=0.95\textwidth,height=0.95\textheight]{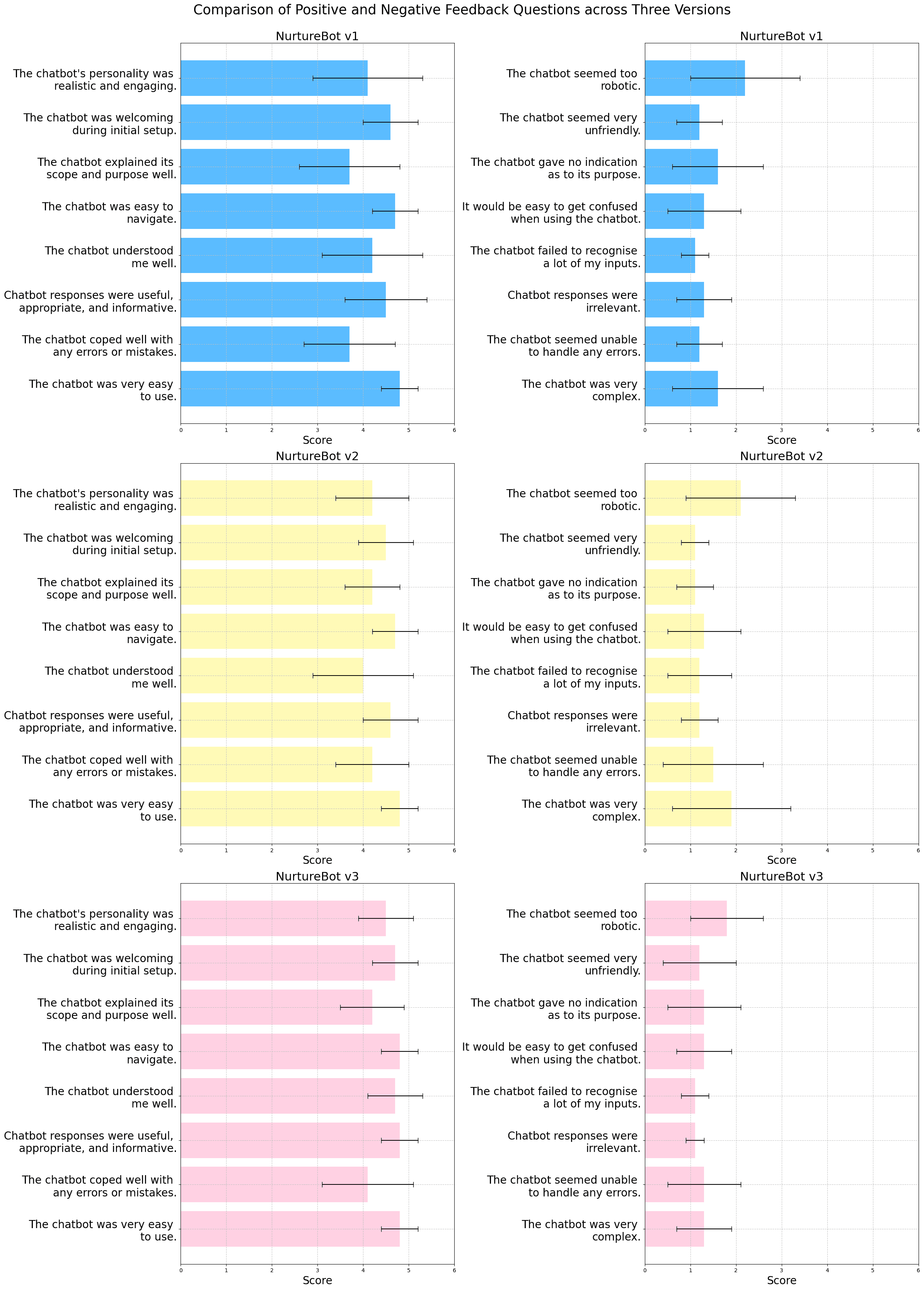}
  \caption{Comparison of CUQ questions of three Versions of NurtureBot. This figure displays the differences between NurtureBot v1, v2, and v3 across various chatbot usability dimensions. The left column contains positive feedback questions, while the right column contains negative feedback question, showing clear improvements across versions, particularly in how engaging and easy to use the chatbot became over time.}
  \Description{A comparison chart showing the usability scores of NurtureBot v1, v2, and v3. The figure contains two sets of bar graphs for each version: one set for positive statements (left) and another for negative statements (right). NurtureBot v1 (blue) shows more neutral scores, while v2 (yellow) and v3 (pink) show improvements in positive responses and a reduction in negative responses. This comparison demonstrates progressive enhancement in user satisfaction across all chatbot versions.}
  \label{cuqq}
\end{figure*}

Further, quantitative results from the \textbf{structured CUQ}  \cite{cuq} indicate a steady improvement in usability across all three versions of NurtureBot. As seen in the CUQ score comparison in Figure \ref{cuq}, the average CUQ score increased from 85.4 in NurtureBot v1 to 87.2 in v2, and further to 91.3 in v3. Notably, the lowest scores also saw significant improvements, rising from 43.8 in v1 to 67.2 in v2, and then to 75 in v3, reflecting enhanced user satisfaction over time. These results confirm that the iterative design process enhanced user experience, making the chatbot more intuitive, engaging, and easier to navigate. From NurtureBot v1 to v3, positive feedback improved significantly, with users rating the chatbot as increasingly engaging, easy to navigate, and better at understanding and supporting their needs (see Figure \ref{cuqq} Left). Negative feedback, such as the chatbot being perceived as "robotic" or "unfriendly," decreased with each version, showing that iterative improvements effectively addressed these issues (see Figure \ref{cuqq} Right). By v3, users reported the most satisfying experience, with fewer errors and a more natural interaction flow, and a decreased overall variance (see Figure \ref{cuq}).

Furthermore, in comparison with related literature at the time of writing this paper, to the best of our knowledge, NurtureBot v3 had achieved the highest reported CUQ score of 91.3 (See Table \ref{tab:chatbots}). While early rule-based systems such as MYA using FlowXo \cite{larbi2022}, and ChatPal using Rasa \cite{boyd2022} reported lower scores when compared to more advanced NurtureBot versions using GPT-4 and DIAS using GPT-4 Turbo \cite{fersch2023}, which achieve significantly higher scores. Notably, despite all NurtureBot versions using the same base technology, the iterative development of a human-centred interaction layer led to significant improvements in usability, demonstrating that thoughtful design can enhance user experience even without changes to the underlying technology.

\begin{table*}[H]
\centering
\caption{In comparision with CUQ scores reported in literature NurtureBot v3 achieves the highest CUQ score of 91.3, followed by NurtureBot v2 and v1, demonstrating progressive improvements in usability, highlighting the effectiveness of iterative human-centred design in enhancing chatbot performance.}
\begin{tabular}{|p{4cm}|p{3cm}|p{3cm}|p{2cm}|p{1cm}|}
\hline
\textbf{Chatbot Name} & \textbf{Base Technology} & \textbf{Domain} & \textbf{Participants} & \textbf{CUQ Score} \\ 
\hline 
MYA\cite{larbi2022} & FlowXo (Rule-based) & Physical Activity  & 30 & 57.4   \\ 
\hline
ChatPal\cite{boyd2022} & Rasa & Mental Health & 10 & 65  \\
\hline
Chat Ella\cite{zhang2024} & GPT-2 & Chronic Diseases & 64 & 68.31  \\ 
\hline
GPT-Simulated Patient\cite{holderried2024} & GPT-3.5 & Medical Education & 28 & 77  \\
\hline
DIAS\cite{fersch2023} & GPT-4 Turbo & Educational Services & 103 & 80  \\
\hline
\textbf{NurtureBot v1} & GPT-4 & Parental Wellbeing & 32 & 85.4 \\
\hline
\textbf{NurtureBot v2} & GPT-4 & Parental Wellbeing & 32 & 87.2 \\
\hline
\textbf{NurtureBot v3} & GPT-4 & Parental Wellbeing & 64 & 91.3 \\
\hline
\end{tabular}

\label{tab:chatbots}
\end{table*}

Based on our findings for co-designing user-LLM interactions, in Parts I-IV, we derive implications for design of AI-driven parental support systems presented in Table \ref{tab:implications}.

\begin{table}[h!]
\centering
\caption{Implications for Design of AI-driven Parental Support Systems}
\begin{tabular}{|p{1.80cm}|p{6.15cm}|}
\hline
\textbf{Implication} & \textbf{Description} \\
\hline
\textbf{Good Data Sources for AI} & A key foundation for successful AI is access to reliable and well-structured data sources. NurtureBot's effectiveness depended on the quality and relevance of the information provided. Future iterations must use up-to-date, verified sources, including health organisation-endorsed health guidelines and localised resources, to enhance trust and provide contextually relevant information. \\
\hline
\textbf{User Understanding, Control, and Outcome Improvement} & Effective human-centred AI balances understanding, controlling, and improving interactions. Users should have transparency in their interactions with the AI and be able to guide the interaction through dynamic in-situ controls. Enabling users to refine the agent’s responses enhances the quality of interactions and makes the AI more responsive to user needs. \\
\hline
\textbf{Personalisation, Contextualisation, and Localisation} & Users expect personalised interactions, asking for the chatbot to remember personal details such as their names, their child’s name, where they live, and their previous conversations. These aspects help the chatbot tailor responses to user’s specific needs. Future designs should consider secure methods for personalisation without compromising privacy. \\
\hline
\textbf{Community and Peer Support} & Users desire more community-oriented support for parental wellbeing. This reflects the need for AI systems to connect users with localised communities, parenting groups, or forums. AI could act as a bridge between users and real-world community resources, providing emotional and practical support. \\
\hline
\textbf{Multi-modal Interactions} & Users expressed a need for multi-modal interactions such as video, voice, and image-based communication alongside text. This allows flexibility in how users interact with the chatbot, broadening accessibility for users with different preferences, literacy levels, or disabilities. \\
\hline
\textbf{Emotional Design} & Emotional connection emerged as a key element of success for NurtureBot. Many participants referred to the assistant as a "friend," indicating the importance of empathy and emotional support. However, it's crucial to ensure users understand the AI’s role as a tool and avoid fostering over-reliance. \\
\hline
\textbf{Co-Design and Iterative Testing} & Co-design played a critical role in aligning NurtureBot’s features with user needs. Iterative testing allowed us to refine the system based on feedback. This iterative process of testing, redesigning, and retesting ensured that the final product provided the best possible user experience. \\
\hline
\end{tabular}
\label{tab:implications}
\end{table}

\begin{figure*}[t]
  \centering
  \includegraphics[width=0.55\textwidth, keepaspectratio=true]{img/PromptArch.pdf}
  \caption{Comparison of prompt architectures across NurtureBot versions, highlighting the co-designed interaction layer introduced in v2 and refined in v3.}
  \Description{xxx}
  \label{promptarch}
\end{figure*}

\section{Discussion}

\sruthi{

\subsection{Co-Design as a Catalyst for LLM-Based Agent Development}
Our study demonstrates the potential of co-design in shaping LLM-based agents, particularly in sensitive domains such as parental wellbeing. Our research aligns with broader HCI research, which shows that empowering users to co-design results in more effective and engaging systems for designing both parental wellbeing \cite{arcmothers,benjamin2024ryelands} and broader AI \cite{hossain2021towards, AIPD, mindset}. On integrating the priorities, lived experiences, and user-generated dialogues directly into the system’s prompt architecture, by designing the interaction layer, we improved the chatbot’s ability to respond empathetically and personally and redefined the benchmarks for usability and user satisfaction in AI-driven systems. 

This approach exemplifies a human-centred AI method, prioritising collaboration and inclusivity in the design process. Our co-design method integrated few-shot dialogue examples generated by participants role-playing as the chatbot directly into the prompt architecture. Unlike traditional rule-based chatbots, which often rely on predefined, inflexible conversational paths, the co-design process capitalised on the generative capabilities of LLMs, leveraging user needs to iteratively refine chatbot interactions, and building on emerging user-centred design methods \cite{benjamin2024ryelands, shi2024wildfeedback, hwang2023aligninglanguagemodelsuser}. This extends existing methods that are typically used for improving LLMs, e.g., fine-tuning, transfer learning, and reinforcement learning from human feedback (RLHF).



The ARC method was instrumental in the co-design process, allowing us to iteratively refine the interaction layer based on user feedback. However, the application of such participatory methods comes with its own limitations. Designing user-aligned systems using ARC requires longitudinal studies and sustained engagement from participants who are deeply invested in the design process. While our implementation successfully captured user needs and incorporated them into the system, scaling such methods may require alternative or complementary design research techniques.

\subsection{The Interaction Layer Prompt Architecture}
Our findings demonstrate that incorporating a user-co-designed interaction layer, employing the tripartite human-centred AI principles of understanding, control, and improving interactions, significantly enhanced the usability and user experience of subsequent versions of NurtureBot. This work contributes to the growing body of research on leveraging human-centred computing for AI alignment \cite{gilbert2024hccneedalignmentthesensible}. The interaction layer prompt architecture (illustrated in Figure \ref{promptarch}), extends techniques like Chain-of-Thought prompting \cite{wei2022chain}, ReAct \cite{yao2022react}, and Deliberative Alignment \cite{guan2024deliberative} to integrates user feedback and employs state transitions to dynamically align system behaviour with human-centred design principles in sensitive contexts.

The Understand, Control, and Improve framework maps closely to Nielsen’s usability heuristics \cite{nielsen}, particularly visibility of system status, user control and freedom, and recognition over recall. These principles address the unique challenges of conversational interfaces, which, unlike modern mobile applications with expansive graphical interfaces, often leave users uncertain about system capabilities. Chatbots can force users into frustrating trial-and-error processes, undermining user trust and engagement. Through our framework, we introduced explicit mechanisms to inform users about the chatbot’s capabilities, guide interactions across states, and empower users to direct conversations toward meaningful outcomes. This approach not only improves usability but also instills confidence in navigating and leveraging the system effectively.

Looking forward, we envision a future where users can seamlessly engage in Understanding, Controlling, and Improving AI systems without the need for rigid frameworks. Transparent and flexible interaction layers hold the potential to enable intuitive, user-aligned engagements that dynamically adapt to diverse needs, ultimately supporting bi-directional human-AI alignment \cite{shen2024bidirectionalhumanaialignmentsystematic}. While this work represents an initial step, we anticipate that advancements in participatory methods and interaction design will further refine and expand the capabilities of generative AI systems, fostering truly human-centred AI experiences.

\subsection{User Anthropomorphism and Ethical Boundaries}
A key insight from our study is that despite our attempts to discourage anthropomorphising the chatbot, participants continued to do so, and even explicitly requested anthropomorphising features through multiple stages of co-design. Participants often described the chatbot as a “friend” or a “companion,” suggesting a deep emotional connection fostered by its empathetic tone and personalised responses. While this can enhance user engagement and perceived support, it raises ethical concerns about over-reliance on non-human systems in emotionally charged contexts \cite{weizenbaum1966eliza}.

These findings reiterate the need for careful balancing between making interactions more empathetic, while maintaining a clear distinction between chatbots and real human relationships.  It will be highly tempting for designers focused on user satisfaction to exploit the ease with which LLM-powered chatbots can mimic human-like empathy. However, the potential for this to lead to parasocial human-AI relationships \cite{maeda2024human} poses huge risks, from the illusion of reciprocity to entrenching social isolation.
Future research should explore how to increase healthy forms of engagement without compromising emotional safety. Novel human-centred AI approaches may be required to counter the ever-present human desire to attribute intelligence and emotion to machines; these could include approaches that encourage contestation \cite{ploug2020four}, centre human choice rather than AI decision-making \cite{miller2023explainableaideadlong}, or simply provide more and better training to help users avoid anthropomorphisation.

}

\sruthi{
\section{Limitations and Future Directions}

Our study offers valuable insights into the co-design of LLM-based parental wellbeing assistants. However, certain constraints warrant consideration, both as reflections on our current findings and as signposts for future inquiry.

\paragraph{Study Design and Population}
Our five-minute daily interaction window sought to ease participant burden, yet this brief engagement may have curtailed the depth and complexity of user responses. Although the data gathered offer a foundational understanding of initial interaction patterns, extended studies would allow richer insights into how user experiences and relationships with the chatbot evolve over time.
We chose to focus on parents of children aged five or six, rather than those navigating the acute perinatal period. While this approach reduced ethical risks--minimising emotional distress and time burdens for parents with newborns--it may have limited the immediacy of the experiences examined. Simulating early stages of parenthood cannot fully capture the depth of challenges faced during pregnancy or postpartum. Future work should explore working with parents currently in these early stages of parenthood, supported by robust ethical frameworks, clinical trials, and collaborations with public health organisations, ensuring that vulnerable populations are engaged responsibly and with adequate safeguards in place.


\paragraph{Risks of Misinformation and Bias}
As with any LLM-based tool, the potential for misinformation or ‘hallucinations’ remains a pressing concern. Although we endeavoured to ensure evidence-based, trustworthy responses, it is not possible to eliminate inaccuracy from LLMs. Integrating domain expert oversight and real-time content verification will be critical to maintaining credibility, especially when the use case could easily spill over into healthcare. Our intention for NurtureBot was to promote parental wellbeing rather than to dispense medical advice; using it for the latter could be highly risky given the poor performance of current LLMs on medical advice tasks \cite{agarwal2024medhalu}.

\paragraph{Saftey}
Safety is a critical concern for a wellbeing agent like NurtureBot. In real-world scenarios, edge cases may emerge where users, either deliberately or unintentionally, prompt the chatbot to generate harmful or inappropriate content. Such cases could lead to emotional distress or misinformation that exacerbates a user's situation. Ensuring that the chatbot does not generate content that could cause harm—whether directly or indirectly—requires robust guardrails, such as toxic content filters, conversational safeguards, and monitoring mechanisms.

\paragraph{Generalisability and Context Sensitivity}
Our findings are anchored in a specific demographic and cultural setting from the United Kingdom. Parenting experiences vary greatly across sociocultural, economic, and personal contexts, and future research should consider more diverse populations. Tailoring LLM-based supports to different communities and circumstances will enhance relevance and ensure broader applicability.

}



\section{Conclusion}
The significance of comprehensive perinatal care cannot be overstated, as it plays a pivotal role in mitigating risks associated with perinatal mind-body duality issues, fostering a positive environment that promotes healthy development of the parent(s), the child, and the long-term stability of the family unit. We explored iterative co-design of AI-powered parental wellbeing support with NurtureBot, engaging users through  ARC sessions, where they were not only able to prioritise key issues but also contribute to the redesign of user-LLM interactions, for improved user comprehension, control, and task outcomes. The resulting interaction layer significantly improved the user experience across versions, suggesting that AI-driven systems can better align with real-world needs by empowering users to shape their design. By prioritising the human experience, AI tools can not only enhance wellbeing but also foster stronger, more resilient communities and create scalable, long-lasting solutions to meet diverse real-world needs.


\begin{acks}
 We express our sincere gratitude to all the parents who participated in our study, with special thanks to the 32 parents who proactively contributed to the co-design of NurtureBot. Your insights and dedication were invaluable in shaping this project. We also extend our heartfelt thanks to the children's organisation, Committee For Children, for generously sharing their wealth of knowledge and resources from the Nurture project. This research is supported by UKRI Future Leaders Fellowship (MR/T041897/1). 

\end{acks}

\bibliographystyle{ACM-Reference-Format}
\bibliography{main}

\appendix
\textbf{APPENDIX}
\section{Qualitative Results from Part I NurtureBot v1 Testing and Analysis}

A detailed account of the seven key problems identified through thematic analysis of participant feedback from the initial trials of NurtureBot v1 in Section \ref{part2}, serving as a foundation for subsequent design iterations.

\textbf{Problem 1: Unable to Chat}: Participants expressed frustration with NurtureBot’s tendency to cut off their conversation. They were disappointed when the chatbot redirected them to external URLs rather than providing concise information within the chat itself. Even when information or exercises were provided within chat, NurtureBot did not engage further, making it difficult to continue the conversation, thereby rendering the participants unable to experience empathetic chatting, which was a key feature. They found this approach disruptive and felt it hindered the conversational flow. Participants noted:
    \begin{quote}
    "The only thing I would maybe find better is instead of providing links to articles, 
    maybe include the information from the articles into the messages from NurtureBot. 
    That way you haven't got to click off from NurtureBot and could spend longer speaking."
    \end{quote}
        \begin{quote}
    "Today I was trying to see if NurtureBot would focus more on chatting and less on giving infomration."
    \end{quote}
    
\textbf{Problem 2: Patronising and Robotic Response Tone}: Participants reported that the NurtureBot’s responses felt patronising and robotic, often pretending to understand human emotions or pain. For example, a participant stated:
    \begin{quote}
    "What's the point of platitudes like `recovering from a 
    c section can be painful' - what does an AI know?? Why would it be helpful to hear that expressed?"
    \end{quote}
    Another participant commented:
    \begin{quote}
    "I think maybe make it slightly more warm and more human-like rather than `here’s your information.'"
    \end{quote}
    
\textbf{Problem 3: Lack of Localised Resources}: The resources suggested by the chatbot were not tailored to participants' specific country or area, making the information less relevant and useful. One participant highlighted:
    \begin{quote}
    "When asking for groups in my area, it needs to be more specific, especially for 
    the area I live in as it came up with places that were a bit irrelevant."
    \end{quote}
    
\textbf{Problem 4: Memory and Continuity in Conversations}: Participants wanted NurtureBot to remember details about them, such as their name, their baby's name, and their troubles. The lack of continuity made the interactions feel impersonal and disconnected. As one participant mentioned:
    \begin{quote}
    "I think it works very well the way it is. Obviously I don't know fully how it works 
    but it would be very useful if it remembers everything that you ever tell it so it 
    builds up a profile of the child and parents so you don't have to keep repeating information."
    \end{quote}

\textbf{Problem 5: Generic responses}: Participants felt that NurtureBot's responses lacked specificity and depth, often resorting to external links without giving user a chance to discuss their situational need and providing direct, tailored answers based on reliable sources. This approach was seen as redundant, as users felt they could achieve similar results using standard search engines. One participant expressed: 
    \begin{quote} 
    "Pointless. I wanted something specific, not links. Don't understand why I wouldn't just use Google." 
    \end{quote}
    
 \textbf{Problem 6: Transactional Nature of Conversations}: The chatbot was perceived as overly transactional, asking questions in rapid succession without allowing participants enough time to process the information or engage in a more natural, flowing conversation. One participant described this experience:
    \begin{quote}
    "It was OK. It still feels quite transactional - NurtureBot wants me to ask questions 
    then I get the answer, and then it asks me the next question. There’s no sense of 
    pausing to process what’s been said, and asking the next question immediately feels 
    a little bit like McDonald’s drive through when you’re asked after every item 
    `is that all?'. It would be nice to feel like the conversation is a bit less like ping pong."
    \end{quote}

 \textbf{Problem 7: Content Overload}: Participants found the amount of information provided in NurtureBot’s responses overwhelming, particularly when the chatbot delivered long paragraphs filled with multiple links. This made it difficult to process the information and maintain a meaningful connection with the conversation. As one participant remarked:
    \begin{quote} "Ok, but it felt like the long responses with lots of links were a bit overwhelming and lacking in connection." 
        \end{quote}

\section{Qualitative Results from Part II ARC Activity Two}

A comprehensive mapping of seven user needs, derived from thematic analysis of dialogues created during participant role-play as NurtureBot in ARC sessions (see Section \ref{part3}), highlighting the co-designed inputs that shaped the interaction layer.

\textbf{User Need 1. Use of Metaphors for Understanding:}
Parents suggested a range of metaphors to help them better understand and form a meaningful connection with NurtureBot. Rather than perceiving the chatbot as a purely functional AI assistant, many participants preferred to imagine NurtureBot as a trusted companion, often referring to it as a ``friendly advice guru.'' This metaphor conveyed the idea of an accessible, approachable figure offering guidance and support in a non-judgemental manner. Others went further, likening NurtureBot to non-interactive entities such as a ``personal diary'' or a ``soundboard,'' highlighting its role as a private space where they could express thoughts and feelings without interruption. These metaphors gave NurtureBot a more intimate, relatable persona, which parents felt made it more approachable and trustworthy. They wanted NurtureBot to introduce itself as a ``trusted friend'' or ``assistant,'' someone they could confide in without fear of judgement, a crucial aspect of the chatbot’s role in supporting parents through the often overwhelming journey of early parenthood.

This metaphorical framing was not merely about making NurtureBot feel more approachable; it also shaped how parents engaged with it. 
For instance, treating the chatbot as a ``friend'' with whom they could converse ``in their own words'' alluded to understanding a more natural, free-flowing interaction, allowing parents to speak candidly about their concerns in natural language. 
The metaphors enable parents to treat NurtureBot as an alternative (digital) companion; one that was aware of their individual queries and specific needs. This blending of technology with a human-like empathy was key to fostering deeper, more effective engagement.

\textbf{User Need 2. Beginning with Empathy:}

Parents highlighted the importance of empathetic conversation starters, expressing that this initial approach sets the tone for the entire interaction. They wanted NurtureBot to start by first asking users how they were feeling or inquiring about their day, creating a welcoming and supportive tone. Parents expressed that this approach was particularly useful in emotional situations, where a simple, ``How are you today?'' can help, offering comfort and reassurance before moving on to problem-solving.

Moreover, empathy needed to extend beyond the initial greeting, especially after they disclosed a personal problem or negative feeling. Rather than immediately diving into solutions or resources, NurtureBot’s ability to show understanding and validation before proposing options was seen as key to making parents feel heard and supported. Whether engaging in mindful exercises or offering parenting information, the empathetic tone had to persist throughout, without pretending to be human or to understand human pain, thereby reinforcing NurtureBot's supportive role.

\textbf{User Need 3. Introducing Features Clearly:}

A recurrent goal of parents was to highlight the importance of clearly introducing NurtureBot’s core features early in the interaction. While participants recognised that a comprehensive list of features was appropriate for the first onboarding session, they also left clear reminders of NurtureBot’s capabilities that should be available throughout subsequent interactions. 

In practice, parents wanted NurtureBot to weave in these features naturally during conversations, reminding them about available options and how to best utilise them in context. This kept the interaction fluid and allowed parents to feel that they could rely on the chatbot’s range of support without needing to recall its functionalities from memory. By continuously reinforcing NurtureBot’s capabilities, within the context of the conversation, it became easier for parents to make informed choices during the interaction, thereby maximising the effectiveness of each conversation.

\textbf{User Need 4. Providing Step-by-Step Guidance:}

Parents wanted NurtureBot to go beyond simply providing information. When guiding them through exercises—whether mindfulness activities, meditations, or stretches—they desired a step-by-step approach that offered a clear, manageable flow of actions. They wanted Nurture to ask when they were ``ready,'' to ``wait for them'' to finish a step before moving on to the next, or even to ``pause'' and ``continue later.'' This guidance needed to be personalised to their situation, transforming NurtureBot into a collaborative tool rather than just a passive source of information. 

Beyond wellbeing exercises, parents also wanted this step-by-step guidance for other interactions, such as when presenting parenting information or helping them manage their schedules. Breaking down long paragraphs of information into bullet points and allowing them to control the flow of the conversation through a step-by-step guide, with parents "choosing the next step" from a range of available options, empowered parents, making them feel in control of the interaction and able to absorb information at their own pace.

\textbf{User Need 5. Personalising and Deepening Sessions:} 

Personalisation was crucial to the parents’ experience with NurtureBot. Parents did not just want generic advice pulled from trusted resources. They wanted the chatbot to understand their specific situation before suggesting exercises or offering advice. Parents had rewritten NurtureBot’s dialogues to either ``browse a list of topics'' or ``dive deeper'' into one specific area depending on their immediate needs. They also wanted NurtureBot to "talk about their situation," for some exchanges before going about finding useful information. This deep personalisation allowed for a more meaningful and tailored exchange, where NurtureBot could address the unique concerns of each parent based on the context of their lives.


\textbf{User Need 6. Offering Clarification and Reassurance:}

At the end of each interaction, parents wanted NurtureBot to provide them with an opportunity to ``clarify'' or ``revisit'' the topic at hand. Whether they had completed an exercise or received parenting advice, participants wanted NurtureBot to ask if they needed further clarifications or wanted to explore additional options. This ongoing dialogue reassured parents that they were in control of the conversation, allowing them to slow down and fully understand the information presented and improve outcomes.

During the course of the conversation, when parents were unsure of what they wanted, participants wanted NurtureBot to ``take the conversation slow,'' offering clarification or suggesting alternative approaches, would ensure that they did not feel rushed or overwhelmed. This reassurance, especially when discussing difficult topics, coupled with the option to fine-tune their experience, would take NurtureBot beyond offering static advice.

\textbf{User Need 7. Summarising and Offering Follow-Up Options:}

Summarisation played a key role in allowing parents to reflect on their interactions and ``takeaways'' with NurtureBot. At the end of each session, parents desired a brief summary of what had been discussed, along with suggestions for follow-up actions. This summary could be ``saved for future reference,'' allowing parents to revisit important information without needing to restart the conversation. Participants also wanted Nurture to offer to ``bookmark, save, or email'' summaries. Additionally, they wanted NurtureBot to offer suggestions for similar exercises, revisit the session, or offer further support if the previous interaction was not sufficient.

Finally, participants wanted NurtureBot to link to buttons or hints to click through various follow-up options, helping parents with choices to guide the conversation forward without needing to articulate their next steps. Imagination of this interactive flow of suggested participants wanted to prevent decision fatigue with more seamless and intuitive interactions, reinforcing the effectiveness of NurtureBot as a comprehensive support tool even for new and non-technical users.








\end{document}